\documentclass{article}
\usepackage{graphicx} 
\usepackage{jheppub}
\usepackage{slashed}
\usepackage[T1]{fontenc}
\usepackage{braket}
\usepackage{multirow}
\usepackage[normalem]{ulem}
\usepackage{graphicx, graphics, hyperref, amsmath, amssymb, slashed, color, bbm,amsthm, array}

\subheader{\hfill \rm KEK-TH-2685}

\title{High Energy Vector Boson Scattering in Four-Body Final States to Probe Higgs Cubic, Quartic, and HEFT interactions.}

\author{Shameran Mahmud$^{1}$ and Kohsaku Tobioka$^{1,2}$}
\affiliation{\vspace{2mm} $^1$ Department of Physics, Florida State University, 77 Chieftan Way, Tallahassee, FL 32306, USA\\
$^2$Theory Center, High Energy Accelerator Research Organization (KEK), 1-1 Oho, Tsukuba, Ibaraki
305-0801, Japan}

\emailAdd{sm19bh@fsu.edu}
\emailAdd{ktobioka@fsu.edu}

\abstract{We compute the energy scales of perturbative unitarity violation in various $2 \to 4$ vector boson scattering (VBS) processes and compare them to lower multiplicity processes. The final states include $h^4$, $V_L V_L h h$, and $V_L^4$, where $V_L$ represents the longitudinal mode of $Z$ or $W$ boson, and $h$ the Higgs boson. We consider scenarios with modified cubic and quartic Higgs self-couplings, including those derived from the Standard Model Effective Field Theory (SMEFT), as well as scenarios involving derivative operators from Higgs Effective Field Theory (HEFT). Modified Higgs self-couplings typically lead to perturbative unitarity violation in at least the $2\to 3$ VBS; however, the corresponding energy scales of unitarity violation are often very high. Our analysis reveals that, in the case of modified Higgs potentials,  $2\to4$ processes exhibit significantly lower energy scales of unitarity violation compared to  $2 \to 3$ processes. This, combined with the fact that the cross sections of $2 \to 4$ processes increase with energy, suggests they generate more signal events at high energies, $\sqrt{\hat{s}} \gtrsim 2$\,TeV, which could be achieved in future colliders. In contrast, for HEFT derivative interactions, higher multiplicity offers diminished benefits, as $2 \to 4$ cross sections are often smaller than those of related $2 \to 3$ processes within the valid energy range. This study shows that $2 \to 4$ VBS processes are particularly compelling for probing Higgs potential modification, including the cubic and quartic couplings, but are less advantageous when derivative interactions dominate.}

\begin{document}
\maketitle
\flushbottom

\section{Introduction}

One of the major questions of particle physics, particularly after the Higgs boson discovery, is whether the Higgs potential aligns with the Standard Model (SM) predictions. Understanding the Higgs potential provides crucial insights into physics beyond the Standard Model (BSM), such as the nature of the electroweak phase transition. The Higgs mass and the vacuum expectation value (VEV) have been measured~\cite{ATLAS:2022vkf, ATLAS:2024fkg, CMS:2022dwd}, allowing for a prediction of the remaining Higgs cubic and quartic couplings within the SM. Therefore one can test potential deviations in these couplings from the SM predictions, as such deviations provide essential clues about the underlying BSM physics.

Measuring the Higgs cubic coupling is a crucial step in exploring the Higgs potential. The primary probe is the di-Higgs production via gluon fusion, $gg \to hh$, where $g$ is a gluon and $h$ a Higgs boson, as studied in Refs.~\cite{Baglio:2012np, Dolan:2012ac, Davies:2019dfy, Chen:2019lzz, Grazzini:2018bsd, Dreyer:2018qbw}. However, the di-Higgs analyses suffer from a low production cross section, resulting in current LHC bounds on the relative coupling strength at $\mathcal{O}(5)$~\cite{ATLAS:2021tyg, ATLAS:2023gzn, CMS:2022gjd, CMS:2022hgz}, {with constraints from perturbative unitarity giving similar results~\cite{DiLuzio:2017tfn}. Bounds on the cubic coupling} are expected to improve to $\mathcal{O}(50\%)$ at the forthcoming HL-LHC~\cite{Cepeda:2019klc}. Given the need for greater precision, developing independent probes to constrain the Higgs cubic coupling is compelling. 

In our previous work~\cite{Mahmud:2024iyn}, we considered high-energy $2 \to 3$ vector boson scattering (VBS), such as $V_L V_L \to V_L V_L h $ and $V_L V_L \to h h h$ where $V_L$ represents the longitudinal mode of $Z/W$ boson. If the Higgs cubic coupling is modified, the $s$-wave amplitude of $2 \to 3$ VBS grows with energy, eventually violating perturbative unitarity. Nevertheless, the cross sections of the $2 \to 3$ VBS remain constant, so energy growth in the cross section could improve precision at proposed future colliders, such as the FCC or a muon collider, where collision energies exceed those of the LHC. Energy-growing cross sections can be realized by VBS at higher multiplicities represented by $V_L V_L \to V_L V_L hh, h^4$, and $V_L^4$. Moreover, previous studies~\cite{Chang:2019vez,Abu-Ajamieh:2020yqi} suggest that the energy scale of unitarity violation for  $V_LV_L\to V_L^4$ is lower than that of the $2 \to 3$ counterparts.

When it comes to the Higgs quartic coupling, the current strategies face several challenges. Unitarity requires that the coupling strength remain below $\mathcal{O}(50)$~\cite{Stylianou:2023xit}. At the HL-LHC, this bound could be improved to $\mathcal{O}(20)$ through analyses of $gg \to hhh$ production~\cite{Papaefstathiou:2023uum, Stylianou:2023xit, Fuks:2017zkg}. Additional studies~\cite{Plehn:2005nk, Bizon:2018syu, Bizon:2024juq, Stylianou:2023xit, Liu:2018peg} include considerations of ILC and FCC setups~\cite{Stylianou:2023xit, Liu:2018peg}. Our approach explores a new probe through high energy VBS. The $V_L V_L \to h h h$ scattering is sensitive to both quartic and cubic couplings, unlike the $V_L V_L h$ and $V_L^4$ final states. This unique sensitivity is highlighted in studies~\cite{Chiesa:2020awd, Stylianou:2023xit}, which indicate that quartic coupling sensitivity could reach $\mathcal{O}(5)$ at future muon colliders via the $\mu^+ \mu^- \to hhh \nu \nu$ process, effectively a form of $V_L V_L \to h h h$. This motivates further VBS studies, particularly $V_L V_L \to V_L V_L h h$ and $V_L V_L \to h^4$, since benefits found in cubic coupling modification, such as energy-growing cross sections and lower energy scale of unitarity violation, also apply to the quartic coupling modification. 

In this work, based on the compelling $2\to4$ VBS processes mentioned,  $V_L V_L \to h^4$, $ V_L V_L hh$, and $V_L^4$, we compute the energy scales of unitarity violation and the cross sections under various modifications of the Higgs potential, comparing them to $2 \to 3$ VBS results to identify the advantages of $2 \to 4$ VBS. Since the focus is the high-energy regime, we apply the equivalence theorem, replacing $V_L$ with the corresponding Nambu-Goldstone (NG) boson. In all cases, we find much lower energy scales of unitarity violation in the $2\to4$ processes. Furthermore, their cross sections exceed those of $2 \to 3$ processes in (parton) energies of a few TeV --- achievable at the FCC and muon colliders. By leveraging the higher energies at future colliders, high energy $2 \to 4$ VBS can be a compelling channel for probing both Higgs cubic and quartic couplings. 
While we have not quantified the backgrounds for either process, $2 \to 4$ scattering could achieve even higher signal significance if it exhibits smaller background yields than $2 \to 3$ scattering, as is often the case. This would give $2\to 4$ VBS an advantage at even lower energies in future colliders. 
Another important feature is that different VBS processes probe the Higgs potential independently, leading to distinct energy growing behavior in the cross sections, as shown in Sec.~\ref{sec:Higgs-potential} and Sec.~\ref{sec:x-sec}. For instance, the $V_L V_L \to V_L V_L hh$ processes are sensitive to the quartic coupling, while the $V_L V_L \to V_L^4$ processes are not, making the former uniquely capable of probing a purely quartic modification. Thus, utilizing different processes to probe both Higgs cubic and quartic couplings provides deeper insights into the underlying BSM theory. Other works, such as Refs.\cite{Falkowski:2019tft, Chang:2019vez, Cohen:2021ucp, Gomez-Ambrosio:2022qsi, Delgado:2023ynh}, also demonstrate energy-growing behavior in multi-Higgs boson processes, while similar studies in final states involving vector bosons include Refs.~\cite{Henning:2018kys, Chen:2021rid, Belyaev:2012bm, Abu-Ajamieh:2020yqi}.

There are several ways to modify the Higgs cubic and quartic coupling. Using the effective field theory (EFT) approach, two such methods are Higgs Effective Field Theory (HEFT)~\cite{Feruglio:1992wf, Bagger:1993zf, Koulovassilopoulos:1993pw} and Standard Model Effective Field Theory (SMEFT)~\cite{Buchmuller:1985jz, Leung:1984ni, PhysRevLett.43.1566}. The key difference between these theories lies in non-decoupling effects, with previous studies distinguishing these frameworks including Refs.~\cite{Alonso:2015fsp, Alonso:2016oah, Cohen:2020xca, Gomez-Ambrosio:2022why}. Notably, SMEFT can be viewed as a subset of HEFT, specifically HEFT without non-decoupling effects. For clarity, we refer to theories with a decoupling limit as SMEFT and those without as HEFT. Throughout this work, we apply both HEFT and SMEFT. Using HEFT, we independently modify the cubic and quartic terms of the Higgs potential, which is not feasible in SMEFT where these terms are correlated. To demonstrate this difference, we also analyze a dimension-six SMEFT operator to examine how its correlations affect cross sections compared to HEFT-like scenarios with independent modifications. In addition, other HEFT operators that modify the Higgs self-couplings are also studied. These non-analytic operators naturally arise in UV models which admit a HEFT, such as those discussed in Ref.~\cite{Banta:2021dek}. 

To expand on the analysis, we also investigate derivative interactions in HEFT, as their high-energy behavior differs from that of modified Higgs potentials. We find similar advantages in $V_LV_L\to h^4$ compared to $V_LV_L\to hhh$, as the cross section of the $2\to 4$ process exceeds that of the $2\to 3$ process at collision energies of a few TeV. However, in final states involving vector bosons, the energy scales of unitarity violation are usually in the same order for both $2\to4$ and $2\to3$ processes. As a result, $V_L V_L \to V_L^4$ and $V_L V_L \to V_L V_L hh$ scattering are less advantageous in these BSM scenarios. Additional studies on HEFT derivative interactions can be found in Refs.~\cite{Belyaev:2012bm, Abu-Ajamieh:2020yqi, Cohen:2021ucp, Kanemura:2021fvp, Garcia-Garcia:2019oig, Davila:2023fkk, Anisha:2024xxc}.

This paper is organized as follows: In Sec.~\ref{sec:HEFTintro}, we go over the differences between SMEFT and HEFT, and show the correlations which arise in the former theory. In Sec.~\ref{sec:PUV}, we review perturbative unitarity violation formalism, and observe how the scale of perturbative unitarity violation changes with respect to the numbers of final state particles. In Sec.~\ref{sec:Higgs-potential}, we present and discuss the results for the Higgs potential modifications for the given $2\to4$ processes. In Sec.~\ref{sec:HEFTderivative}, we repeat the computation for a few HEFT derivative modifications. In Sec.~\ref{sec:x-sec}, we compare the cross sections of the $2\to4$ processes with that of the $2\to 3$ processes for both the potential and derivative modifications.

\section{Comparing HEFT to SMEFT}\label{sec:HEFTintro}

We examine the energy scales of perturbative unitarity violation in HEFT as well as SMEFT. The parametrization and differences between the EFTs are briefly explained in this section.

\subsection{EFT Parametrizations}
At a glance, the main difference between SMEFT and HEFT is simply the parametrization of the Higgs sector in the Lagrangian. In the SMEFT, the Higgs field is parametrized as the $SU(2)$ Higgs doublet. For example, using a linear parametrization,

\begin{equation}
\label{eq:su2}
    H = \frac{1}{\sqrt{2}}\left( \begin{array}c  \sqrt{2}G^+ \\ (v+h)+iG^0 \end{array}  \right),
\end{equation}
where $h$ is the singlet Higgs field, $v= 246$ GeV is the Higgs VEV, and $G^+$ and $G^0$ are the NG bosons associated with the spontaneous symmetry breaking. Now, new terms in the Higgs potential are written as integer powers of $H^\dag H$,
\begin{equation}
\label{eq:smeft}
    V_{\rm SMEFT} = \sum_n \frac{c_{2n}}{\Lambda^{2(n-2)}} \left(H^\dag H \right)^n.
\end{equation}
where $c_{2n}$ is a dimensionless parameter and $\Lambda$ is the mass scale. In the HEFT case, the Higgs potential can be written in the unitary gauge by using the singlet, $h$, from Eq.~\eqref{eq:su2},  
\begin{equation}
\label{eq:heft}
    V_{\rm HEFT} = \sum_\ell \tilde{c}_\ell h^{\ell},
\end{equation}
where the $\tilde{c}_\ell$ are dimensionful coefficients.  Eq.~\eqref{eq:smeft} can also be expanded in the unitary gauge, as $H^\dag H = \frac{(v+h)^2}{2}$,  and then the $c_{2n}$ can be matched to  $\tilde{c_\ell}$, with correlations between different $\tilde{c_\ell}$'s. 

One benefit of using the singlet, $h$, is that potential terms which are non-analytic as a function of $H^\dag H$ at $H=0$ can be expressed by expanding around $v$, such as
\begin{equation}
    V= g \left( H^\dag H \right)^\frac{2}{3} \supset  g v^\frac{4}{3} \sum_{n=3} \left( \begin{array}c \frac{4}{3} \\ n \end{array}  \right) \left( \frac{h}{v}\right)^n \ . 
\end{equation}
These two examples show the generality of  HEFT compared to SMEFT. This aspect is explored more in Sec.~\ref{subsec:correlations}. And, see Refs.~\cite{Alonso:2016oah, Falkowski:2019tft, Chang:2019vez, Cohen:2020xca} for the discussion of non-analytic nature of HEFT. 

A non-trivial aspect of identifying the HEFT has to do with field redefinitions. The SMEFT (or even the SM) can look HEFT-like by redefining the singlet, $h$, in a non-linear way. The solution is discussed in-depth in Refs.~\cite{Alonso:2015fsp, Alonso:2016oah, Cohen:2020xca, Gomez-Ambrosio:2022qsi}, which involves looking at the geometry of the Higgs sector.

\subsection{Correlations in SMEFT}
\label{subsec:correlations}
In the case of the cubic and/or  quartic coupling shifts, either a HEFT or a SMEFT operator may be admitted depending on the relation between $\delta_3$ and $\delta_4$ in the potential modification, $\delta V(h) = \frac{m_h^2}{2 v} \delta_3 h^3 + \frac{m_h^2}{8 v^2} \delta_4 h^4$. To see this, let us expand Eq.~\eqref{eq:smeft} in the unitary gauge, redefine the mass and VEV to account for the new linear and quadratic terms, and begin matching the coefficients with those in Eq.~\eqref{eq:heft}. Explicitly writing Eq.~\eqref{eq:heft} with the SM potential deviations yields
\begin{equation}
\label{eq:bsm}
    V(h)= \frac{1}{2} m_h^2 h^2+ \frac{m_h^2}{2 v} (1+\delta_3)h^3+ \frac{m_h^2}{8v^2}(1+\delta_4) h^4+\sum_{\ell=5}\frac{\delta_\ell}{v^{\ell-4}} h^\ell, 
\end{equation}
where $\delta_\ell$ are (dimensionless) deviations from the SM. In the case of HEFT, each deviation can generically be independent of one another. However, if a SMEFT operator is of interest, these $\delta_\ell$ must be related to $c_{2n}$. For the first example, consider adding the dimension-six operator $(H^\dag H)^3$ to the Lagrangian. Then Eq.~\eqref{eq:bsm} is truncated at $\ell=6$ and has up to four non-zero $\delta_i$, where $i=3, 4, 5, 6$. $\delta_6$ can be related to $c_6$ by looking at the $h^6$ terms in both Eq.~\eqref{eq:smeft} and Eq.~\eqref{eq:bsm}, giving
\begin{equation}
\label{eq:lambda}
    \delta_6 =  \frac{c_6 v^2}{ 8 \Lambda^2}.
\end{equation}

 However, $c_6$ is the only free parameter that is introduced, as only one operator is added. This means all the other $\delta_i$ must be correlated to $c_6$. Let us choose $\delta_3$ as a free parameter for convenience. Then, by expanding $(H^\dag H)^3$ and matching powers of $h$, the other parameters are expressed in terms of $\delta_3$ as
\begin{subequations}
\label{eq:smeftcorrelation}
    \begin{align}
        \delta_4 &= 6 \delta_3, \\
        \delta_5&= \frac{3 m_h^2}{8v^2} \delta_3, \\
        \delta_6&= \frac{m_h^2}{16 v^2} \delta_3,
    \end{align}
\end{subequations}
making the correlations manifest. So, knowing the value of the Higgs cubic coupling modification predicts the value of the quartic coupling modification; from Eq.~\eqref{eq:smeftcorrelation}, having processes to constrain the quartic coupling more effectively could falsify the dimension-six SMEFT case. This is also discussed in Ref.~\cite{Gomez-Ambrosio:2022why}. 

For the next example, consider adding both the dimension-six operator $(H^\dag H)^3$ and the dimension-eight operator, $(H^\dag H)^4$. Now, Eq.~\eqref{eq:bsm} is truncated at $\ell=8$ and has up to six $\delta_i$, from $i=3$ to $8$. This time, two new operators were added, so there are two potential free parameters, which are chosen to be $\delta_3$ and $\delta_4$. The other parameters are expressed in terms of $\delta_3$ and $\delta_4$ as
\begin{subequations}
\label{eq:dimensioneight}
    \begin{align}
        \delta_5 &= \frac{m_h^2}{8 v^2} (2\delta_4-9 \delta_3), \\
        \delta_6 &=  \frac{m_h^2}{16 v^2} ( 3\delta_4- 17 \delta_3),\\
        \delta_7 &= \frac{m_h^2}{16 v^2} ( \delta_4- 6\delta_3), \\
        \delta_8 &=\frac{m_h^2}{128 v^2} ( \delta_4- 6\delta_3).
    \end{align}
\end{subequations}
 Substituting the condition of only the dimension-six operator, i.e. $\delta_4=6 \delta_3$ returns Eq.~\eqref{eq:smeftcorrelation}.

Note that a HEFT operator can still create correlations in the higher power $h$ terms in Eq.~\eqref{eq:bsm}. As an example, a potential modification such as $\frac{y^4}{4 \pi^2}(H^\dag H)^2 \ln(H^\dag H)$ will create correlations between all powers of $h^\ell$, $\ell \geq 3$, from the term proportional to $\ln(1+\frac{h}{v})$, but with different relations compared to the SMEFT cases in Eqs.~(\ref{eq:smeftcorrelation}, \ref{eq:dimensioneight}). In addition, there will be infinite $\delta_\ell h^\ell$ terms, leading to no finite dimensional truncation for this HEFT operator.


\subsection{Motivating HEFT}

While unitarity violation exists in both  SMEFT and HEFT, there are a few qualities of the latter which make exploring HEFT terms in the Lagrangian compelling as well. 

One of the main reasons of testing  HEFT is that the scale of unitarity violation is low enough to be tested.  HEFT will have unitarity violation at an energy scale of a few TeV, especially as the multiplicity of the final states are increased, as shown in \cite{Falkowski:2019tft, Cohen:2021ucp}. This arises from non-decoupling effects of the new degree of freedom added to the Lagrangian. For this reason, HEFT is more testable and potentially even falsifiable, which would answer the question of whether or not the $SU(2)_L \times U(1)_Y$ gauge symmetry is linearly realized.

The second reason for applying HEFT is that if  the Higgs self-couplings are generically modified, the new potential is more likely to be from HEFT than SMEFT. For example, as shown in Sec.~\Ref{subsec:correlations}, adding a power of $(H^\dag H)^n$ and expanding in terms of $h$ necessarily leads to multiple correlated coefficients. Hence, modifying just one or two terms of $h^\ell$ in Eq.~\eqref{eq:bsm} leads to the HEFT case. If we consider a theory where $\delta_3=0$ but $\delta_4\neq 0$, using the SMEFT potential as described in Sec.~\ref{subsec:correlations}, this scenario requires the coefficients of the dimension-six and dimension-eight operators to be of the same order. Achieving this is challenging; however, it can be realized in HEFT where no correlations enforce relationships between terms. The dimension-six and dimension-eight SMEFT operators will still necessarily turn on the deviations in $h^\ell$ from $\ell= 5$ to $8$ with the correlations in Eq.~\eqref{eq:dimensioneight}, unlike in a general HEFT. For this reason, HEFT is more suitable for a bottom-up approach, as it provides a more general EFT.

A third motivation for HEFT is that a generic HEFT potential gives diverse unitarity violating VBS processes. As an example, let us look at a cubic coupling modification. Shifting just this term in the potential creates non-analyticities. Consider rewriting $h^3$ in a gauge invariant manner,
\begin{align}
\label{eq:h3}
 h^3 \to & \left( \sqrt{2 H^\dag H}- v\right)^3
  \supset h^3 + \frac{3 G^2}{2 v} h^2 \sum_{n=0}^\infty \left(- \frac{h}{v} \right)^n
  + \frac{3 G^4}{8 v^3} \sum_{n=0}^\infty (n+1) \left(- \frac{h}{v} \right)^n \left(2 v h- \frac{n+2}{2} h^2  \right)  \\\nonumber & +\frac{G^6}{16 v^5}  \sum_{n=0}^\infty \frac{(n+1)(n+2)}{2} \left(- \frac{h}{v} \right)^n \left\{ 2 v^2- (n+3)\left( 2vh - \frac{n+4}{4} h^2 \right) \right\} + \cdots 
\end{align}
where $G^2=2 G^+ G^-+(G^0)^{2}$. This modification of the cubic coupling gives an infinite number of Higgs-NG boson contact interactions, whereas a cubic coupling modification in SMEFT, e.g. using $(H^\dag H)^3$, would only have at most, 6-point contact interactions. While each power of $h^\ell$ in $(H^\dag H)^3$ ($\ell=3$ to $6$) creates inifinite contact terms similar to Eq.~\eqref{eq:h3}, most of the terms are canceled after summing them up, thus predicting only a finite number of contact interactions. Other HEFT operators, such as $(H^\dag H)^2 \ln(H^\dag H)$, also have infinite contact terms, without cancellations occuring as in the SMEFT case. We show in Sec.~\ref{sec:PUV} that higher multiplicity processes will lead to stronger energy growth in the $s$-wave amplitude for the $2\to n$ process, starting at $n=3$, regardless of whether a contact term arises from a HEFT or a SMEFT operator. However, as also discussed in Sec.~\ref{sec:PUV}, and as mentioned in Refs.~\cite{Chang:2019vez, Abu-Ajamieh:2020yqi, Falkowski:2019tft}, higher multiplicity processes beyond the $n=3$ case will lead to even stronger energy growth in the $s$-wave amplitude, which will cause it to violate perturbative unitarity at lower energies until reaching final state multiplicities of $n = \mathcal{O}(10)$ in the $2\to n$ process. 

The decrease in energy scales of perturbative unitarity violation is the most apparent when comparing the $2\to3$ process with the $2\to 4$ process. From Eq.~\eqref{eq:h3}, the minimal case to see perturbative unitarity violation is the $2\to 3$ case, as shown in Ref.~\cite{Mahmud:2024iyn}, where we studied perturbative unitarity violation in multiple $2\to3$ VBS processes. While collider searches prefer lower multiplicity states because the analysis is simpler, the stronger energy-growing behavior of higher multiplicities makes it compelling to search for the next-to-minimal case, which would be the $2\to 4$ processes. The benefits of the $2\to4$ processes compared to the $2\to3$ processes are discussed in the next section, Sec.~\Ref{sec:PUV}, as well as in Sec.~\ref{sec:x-sec}.

\section{Energy Scale of Perturbative Unitarity Violation}\label{sec:PUV}

The standard model relies on the Higgs sector for delicate cancellations to control the energy growth of the $W_L W_L \to W_L W_L$ process. If only the diagrams involving the $W_L$ are included, this process leads to energy growth and (perturbative) unitarity violation, which is remedied by including the Higgs boson diagram, which leads to cancellations, saving the perturbative unitarity.

Even in the presence of the Higgs boson, a new EFT term to the Higgs sector breaks similar delicate cancellations, and the unitarity breaks down again at some energy scale, as the cancellations realized in the SM become incomplete. We analyze tree-level perturbative unitarity violation of $2\to 3,4$ VBS to find the energy scale at which the theory breaks down, $E_*$. We use the equivalence theorem to approximate $V_L$ to be the corresponding NG boson because  $E_*$ is high enough to justify the approximation. The main benefit of using the equivalence theorem is that energy-growing behavior is manifest without needing to track cancellations in detail.

We use the formalism of Refs.~\cite{Falkowski:2019tft,Chang:2019vez, Cohen:2021ucp} to obtain the phase-space averaged matrix element, which extracts the $s$-wave part of the matrix element, $\cal M$, normalized to be dimensionless
\begin{equation}
\label{eq:Mavg}
\hat{\mathcal{M}} =  \frac{1}{\sqrt{ S_{\mathrm{in},m} \cdot S_{ \mathrm{out},n}}}\left( \frac{1}{ \int \mathrm{dLIPS}_m \int \mathrm{dLIPS}_n}\right)^\frac{1}{2} \int \mathrm{dLIPS}_m \int \mathrm{dLIPS}_n \mathcal{M}.
\end{equation}
 Here, $S_{\mathrm{in}, m} = \prod_i m_i!$ and  $S_{\mathrm{out},n}=\prod_i n_i!$ are the symmetry factors for incoming and outgoing particles respectively; $m_i$ ($n_i$) are indistinguishable particles in the initial (final) state, $m$ ($n$) is the number of incoming (outgoing) particles,  such that $\sum_i m_i=m$ ($\sum_i n_i=n$), and
 \begin{equation}
  \mathrm{dLIPS}_n = \prod_{i=1}^n \frac{\mathrm{d}^3p_i}{(2 \pi)^3 2 E_i}
 (2 \pi)^4 \delta^4\left(P-\sum_i p_i\right)
\end{equation}
is the Lorentz invariant phase space integral .

To compute $E_*$ and find the strength of tree-level unitarity violation, and hence, sensitivity to new physics, we check the condition,
\begin{equation}
\label{eq:mhatcondition}
|\hat{\mathcal{M}}|^2 \simeq 1.
\end{equation}
The $E_*$ value is related to some new physics scale that is responsible for the modifications in the Higgs sector, but not necessarily the same as the new physics scale. New physics could repair the unitarity before reaching $E_*$, for example, if a new particle can be produced at energies lower than $E_*$. Therefore, we emphasize that $E_*$ is the \textit{maximum} scale at which the unitarity violation must be cured. Also, different scattering processes have different values of $E_*$, and processes with lower $E_*$ values will be more sensitive to new physics in the Higgs sector. 

The derivation of the phase-space averaged matrix element and the connection to unitarity violation are discussed in App. A of Ref.~\cite{Mahmud:2024iyn}. For a constant $\mathcal{M}$, the relation between the cross section, $\sigma$, $\hat{\mathcal{M}}$, and $E_*$ for the $2\to n$ scattering is
\begin{align}
    \sigma \sim \frac{|\hat{\mathcal{M}}|^2}{E_{\rm cm}^2} \sim \frac{1}{E_{\rm cm}^2}\left(\frac{E_{\rm cm} }{E_*}\right)^{2(n-2)},\label{eq:x-sec_scaling}
\end{align}
where $E_{\rm cm}=\sqrt{\hat s}$ is the center-of-mass energy of VBS. This is discussed in more detail in Sec.~\ref{sec:x-sec}.

\subsection{Effect of High Multiplicity on the Unitarity Violation Scale}

Let us now address the enhanced energy growth observed in higher multiplicity states, specifically explaining why the $2\to4$ VBS generically exhibits stronger energy-growing amplitudes than the $2\to3$ VBS. We restrict our analysis to modifications of the Higgs potential only.

Firstly, we consider a $(n+2)$-point contact interaction between a set of scalars, which contributes to $\mathcal{M}$ of the $2\to n$ process. This contribution is independent of $E_{\mathrm{cm}}$, or any other momenta or angles. Other diagrams involving propagators necessarily have a lower power in $E_{\mathrm{cm}}$ than one from the contact term, so they are subdominant at high energy.  Therefore, only the constant matrix element is retained, and all other diagrams involving propagators are neglected. 

Next, we compute  $\hat{\mathcal{M}}$. As the energy scales of interest are much higher than the mass scales of the particles, the massless limit is taken. Since only the contact term is retained in $\mathcal{M}$, we can factorize the amplitude and independently evaluate the $\int \mathrm{dLIPS}_n$ component of Eq.~\eqref{eq:Mavg}. In the high-energy limit, this yields

\begin{equation}
    \int \mathrm{dLIPS}_n = \int \prod_{i=1}^n \frac{\mathrm{d}^3p_i}{(2 \pi)^3 2 E_i}
 (2 \pi)^4 \delta^4\left(P-\sum_i p_i\right) \simeq \frac{1}{8 \pi \left(n-1\right)! \left( n-2 \right)!} \left(\frac{E_{\rm cm}}{4 \pi} \right)^{2(n-2)}.
\label{eq:LIPSn}
\end{equation}
In Eq.~\eqref{eq:LIPSn},  there is always energy growth as long as $n \geq 3$. We give explicit forms of $n=3,4$ cases,
	\begin{align}
    \label{eq:lipscases}
	 \int \mathrm{dLIPS}_3=\frac{E_{\mathrm{cm}}^2}{256 \pi^3},  \quad &\qquad\int \mathrm{dLIPS}_4= \frac{E_{\mathrm{cm}}^4}{24576 \pi^5}.
	\end{align}
There is a much higher power of energy in the $n=4$ case, leading to $\hat{\mathcal{M}}$ having much stronger energy growth compared to the $n=3$ case. However, while the energy growth of the $2\to4$ process is higher, the problem of a larger phase space factor suppressing this energy growth is also possible. To see how the $2\to 4$ process has a lower $E_*$ than the $2\to 3$ counterpart assuming similarly sized potential deviations, one can compute $E_*$ in the most general potential modification case by schematically parametrizing the general $2\to n$ process matrix element as 

\begin{equation}
\label{eq:generalM}
\mathcal{M} = C_n \frac{ \alpha_n \delta}{\Lambda^{(n-2)}},
\end{equation}
where $C_n$ is the combinatoric factor of the diagram, $\Lambda$ is an energy scale obtained from the potential, and  $\alpha_n$ is a numerical prefactor. The assumption that all numerical prefactors are the same for different $n$ is made, i.e. $\alpha_n \equiv \alpha$. While this is not always the case, let us assume this holds for simplicity. 

As an example, consider the cubic coupling modification in Eq.~\eqref{eq:h3} leading to $\Lambda=v$. In the case of $V_L V_L \to h^n$ scattering, the numerical factors are $\alpha= {3 m_h^2}/{(2 v^2)}$ and $C_n= n!$. Note that, in the final states involving $V_L$, the numerical factors are not as simple because different terms contribute to the matrix element. However, we will see this difference will not change the fact that $E_{*, 2 \to 4}$, the unitarity violation scale for the $2\to4$ process, is much smaller than $E_{*, 2 \to 3}$, the unitarity violation scale for the $2\to3$ process.
  
Now, let us consider the special case of the $W_L W_L \to h^n$ process. Using the fact that now, $ S_{ \mathrm{out},n} = n!$, $|\hat{\cal M}|^2$ is written as
\begin{equation}
\label{eq:generalMhat}
|\hat{\mathcal{M}}|^2= \left( \frac{ \alpha \delta}{8 \pi} \right)^2  \frac{n!}{(n-1)! (n-2)!} \left( \frac{E_{\rm cm}}{4 \pi v} \right)^{2(n-2)}. 
\end{equation}
Now, solving for $E_{*, 2 \to n}$, the general unitarity violation scale for the $W_L W_L\to h^n$ process, gives
\begin{equation}
\label{eq:estarn}
E_{*,2\to n} = \left( \frac{8 \pi}{\alpha \delta}\right)^\frac{1}{n-2} \left( \frac{(n-1)! (n-2)!}{n!} \right)^\frac{1}{2(n-2)} \times 4 \pi v.
\end{equation}

Using the gamma function, and making the assumption that $\delta$ is ${\cal O}(1)$, Eq.~\eqref{eq:estarn} is numerically minimized to obtain that $E_{*, 2 \to n}$ is the smallest for $n=10$ in the case of $W_L W_L \to h^n$, which is the same order as Ref.~\cite{Chang:2019vez}, where the $Z_L h^6 \to Z_L h^6$ process was found to have the lowest $E_*$ value. However, the rate of the decrease in $E_*$ is not constant. In the case of $n=3$ and $n=4$, substituting the values in Eq.~\eqref{eq:estarn} shows that 
\begin{equation}
\label{eq:3v4}
    E_{*, 2 \to 3}\sim 10 \cdot E_{*, 2 \to 4},
\end{equation}
while for $n=4$ and $n=5$,
\begin{equation}
\label{eq:4v5}
    E_{*, 2 \to 4}\sim 2 \cdot E_{*, 2 \to 5},
\end{equation}
showing that higher multiplicities will reduce $E_*$ until reaching multiplicities of $\cal O$$(10)$. However, in the case of an $\mathcal O(1)$ deviation, going to higher multiplicities after $n=4$ has much smaller returns than going to the $n=4$ case from $n=3$. 

 While in the case of the final states involving vector bosons, this analysis does not hold exactly, as the symmetry factors and $\alpha_n$ may be different, the symmetry factor multiplied by the different prefactors should be in a similar order as in the Higgs-only final states. This is evident in the fact that $E_{*, 2 \to 3}\sim 10 \cdot E_{*, 2 \to 4}$ even in the case of the final states involving vector bosons, as seen in Fig.~\ref{fig:res1} in Sec.~\ref{sec:Higgs-potential}.

The lower $E_*$ value benefits the cross section. As shown in Eq.~\eqref{eq:x-sec_scaling}, the cross sections scale as $E_{\rm cm}^{2(n-3)}/{E_*^{2(n-2)}}$. 
Although $E_{\rm cm}/{E_*}$ is always a suppression factor in each process, the cross section for the $2\to 4$ process can exceed that of the $2\to 3$ process at sufficiently high energies because of the significantly lower $E_*$ value. We discuss in Sec.~\ref{sec:x-sec} the energy at which the $2\to 4$ cross sections become larger than those of the $2\to 3$ processes.

The situation is more complex for the HEFTs involving derivatives. In contrast to the potential modification case, the matrix element generally depends on angles and momenta, preventing the factorization to compute $\hat{\mathcal{M}}$. The additional integrand in the phase space integral leads to an overall stronger suppression, so that $E_{*, 2 \to 3} \sim E_{*, 2 \to 4}$. Due to the suppression factor of $E_{\rm cm}/E_*$, this means the cross section for the $2 \to 4$ process never gets larger than that of the $2\to 3$ process before the EFT becomes invalid. However, in some special cases, the factorization of $\cal M$ from $\mathrm{dLIPS}_n$ is still possible, and benefits similar to those of the potential modifications hold in those scenarios. The specifics of the HEFT derivative operators are discussed in Sec.~\ref{sec:HEFTderivative} and Sec.~\ref{subsec:derxsec}.

\section{Energy Growth in Higgs Potential Modifications}\label{sec:Higgs-potential}

Based on the modified Higgs potential, we compute $E_*$ values of multiple $2\to4$ processes, $V_L V_L \to V_L^4$, $V_L V_L h h$, and $h^4$. This is a bottom-up approach, as we neglect any derivative terms which can be generated from UV models. The derivative terms are covered in Sec~\ref{sec:HEFTderivative}. We focus on the simplest potential modifications in  HEFT and  SMEFT and several non-trivial HEFT potentials. 

\subsection{Modified Higgs Cubic and Quartic Interactions}
At first, we look at the Higgs cubic modification and quartic modification separately, 

\begin{align}
\delta V=  \frac{m_h^2}{2 v} \delta_3 h^3 + \frac{m_h^2}{8 v^2} \delta_4 h^4\ . 
\end{align}

The cubic coupling modification was also studied in Refs.~\cite{Falkowski:2019tft, Chang:2019vez} in the context of $V_L V_L \to h^n$ processes. Ref.~\cite{Chang:2019vez} also included some vector boson final states, which have similar $E_*$ values as the $2 \to 4$ VBS processes studied here. As a reminder, turning on only the cubic or only the quartic coupling modification while keeping higher order $\delta_{n>4}$ off would admit a HEFT, rather than a SMEFT, as discussed in Sec.~\ref{subsec:correlations}.

For the general polynomial Higgs potential, we compute $E_*$ values as a function of $\delta_i$, and apply it to multiple scenarios. While most of the processes only depend on $\delta_3$ and $\delta_4$, the $h^4$ final state also has a dependence on $\delta_5$, which is defined as $\delta V\supset \delta_5 h^5/v$, as seen in Eq.~\eqref{eq:bsm}. However, as $\delta_5$ has no experimental constraints, it is turned off, unless the new potential modification predicts its value, as is the case for the SMEFT operator $(H^\dag H)^3$, as well as all the HEFT operators inspired by UV models. The dependence on $\delta_i$ in the $E_*$ values for the VBS processes initiated by $W_LW_L $, shown in Fig.~\ref{fig:res1}, are as follows,
\begin{subequations}
\label{eq:estarshifts}
\begin{align}
     E_*(W_L W_L \to hhh) &= \frac{120 \mbox{ TeV}}{|\delta_3-\frac{1}{3} \delta_4|}, \\
    E_*(W_L W_L \to W_L W_L h) &= \frac{71 \mbox{ TeV}}{|\delta_3|}, 
    \\
    E_*(W_L W_L \to h^4) &= \frac{21 \mbox{ TeV}}{|\delta_3-\frac{1}{3} \delta_4+ \frac{10 v^2}{3 m_h^2} \delta_5|^\frac{1}{2}}, \\
    E_*(W_L W_L \to W_L W_L h h) &= \frac{12 \mbox{ TeV}}{|\delta_3-\frac{1}{5} \delta_4|^\frac{1}{2}}, \\
   E_*(W_L W_L \to W_L^4) &= \frac{19 \mbox{ TeV}}{|\delta_3|^\frac{1}{2}},
\end{align}
\end{subequations}
where we include the $2\to 3$ processes for comparison. None of these processes have any dependence on $\delta_{n\geq 6}$. Other $E_*$ values for different processes in both the 3-body and 4-body final states are given in Apps.~(\ref{app:values24}, \ref{app:values23}).

{Note that Refs.~\cite{Chang:2019vez, Abu-Ajamieh:2020yqi} included certain $3\to3$ processes which are analogous to the $2\to4$ processes studied in this paper, such as the $Z_L Z_L Z_L \to Z_L Z_L Z_L$ process. The $E_*$ values would differ due to different phase space and symmetry factors in Eq.~\eqref{eq:Mavg}. For example, comparing $Z_L Z_L \to Z_L^4$ to the $Z_L Z_L Z_L \to Z_L Z_L Z_L$ process, the symmetry factors for the former process is $2\cdot4!$, while the latter process obtains $(3!)^2$. In addition, the phase space integration of $3\to 3$ process is $(\mathrm{dLIPS}_3)^2$ which is different from the one in  the $2\to 4$ process, $ \mathrm{dLIPS}_2 \times \mathrm{dLIPS}_4 $}.

\subsection{HEFT Potentials Motivated by UV models}
\label{subsec:HEFTpotential}

Other potential terms are based on the potentials motivated by several UV models which admit the HEFT. Being HEFTs, all of these terms are non-analytic around $H=0$. Custodial symmetry is assumed for simplicity, as the HEFT can violate custodial symmetry in general. These potentials are
\begin{align}
\label{eq:pot1}
        \delta V&= \frac{y^4}{4 \pi^2} \left( H^\dag H\right)^2  \ln \left( H^\dag H \right), \\
        \label{eq:pot2}
   \delta V&=  \kappa_{\frac{2}{3}} \left( H^\dag H\right)^\frac{2}{3}, \\
   \label{eq:pot3}
    \delta V&=  \kappa_{\frac{1}{2}}\sqrt{H^\dag H}.  
\end{align}
The first potential, Eq.~\eqref{eq:pot1}, arises when integrating out a heavy particle at the loop level. For example, this is seen in Refs.~\cite{Cohen:2020xca, Cohen:2021ucp, Banta:2021dek}, when heavy fermions are integrated out. We also discuss this in more detail in Sec.~\ref{sec:HEFTderivative}. We set $y=1$, and the $4 \pi^2$ factor is left, to emphasize that this potential is loop level. Integrating heavy scalars gives a similar term, but with a different sign and factor. The other two potentials, Eqs.~(\ref{eq:pot2},~\ref{eq:pot3}),  arise in different parameter spaces of the two-Higgs doublet model at tree level. {The $\left( H^\dag H\right)^\frac{2}{3}$ interaction is mentioned in Ref.~\cite{Falkowski:2019tft}}. {We choose  $\kappa_{2/3} = (80$ GeV $)^{8/3}$ based on the bounds for the UV model in Eq.~\eqref{eq:fullsigma}. The corresponding cubic modification given by this choice of $\kappa_{2/3}$ is $\delta_3\sim -0.1$.} As integrating out a heavy field also includes non-analytic derivative terms, we will check those as well in Sec.~\ref{sec:HEFTderivative}. {The $\sqrt{H^\dag H}$ interaction is derived in Ref.~\cite{Galloway:2013dma}, and we choose the scale of the coefficient $\kappa_{1/2}$ to be the same as the scale of $\kappa_{2/3}$ for simplicity. The corresponding cubic modification is $\delta_3 \sim -0.3$, which is not constrained by the Higgs cubic bounds. }

In order to compute the matrix element from a contact interaction, we expand the non-analytic potential up to the needed number of NG bosons in the linear parametrization, as in Eq.~\eqref{eq:su2}. As a schematic example, consider the logarithm potential of Eq.~\eqref{eq:pot1},

\begin{align}
\label{eq:log}
\left( H^\dag H \right)^2 \ln\left( H^\dag H \right) \supset & \frac{v^4}{2} \ln\left( 1+ \frac{h}{v} \right) + \frac{v^4}{4}\ln\left(1+\frac{G^2}{\left(v+h \right)^2}\right) \supset v^2\frac{G^2}{4} \sum_{n=0}^\infty \left( n+1 \right) \left(-\frac{h}{v}\right)^n + \\  &
\frac{G^4}{8} \sum_{n=0}^\infty \frac{\left( n+1 \right) \left( n+2 \right) \left( n+3 \right)}{6} \left(-\frac{h}{v}\right)^n + \cdots \nonumber
\end{align}
 The existence of $\ln\left( 1+ \frac{h}{v} \right)$ means there will be $\delta_5$ terms in the Lagrangian (as well as $\delta_6, \delta_7...$). However, the correlations do not necessarily match those of the SMEFT cases. For our case, we expand up to $\mathcal{O}(G^6)$, $\mathcal{O}(G^4 h^2)$, and $\mathcal{O}(G^2h^4)$ as the $2\to 4$ VBS is the main concern.

Next, after the matrix element evaluation, we average over the momentum using Eq.~\eqref{eq:Mavg}. The massless limit was taken in all cases, as the $E_*$ values are still of $\mathcal{O}(10$)TeV.

\subsection{Results}

\begin{figure}[!tbp]
\centering
\includegraphics[width=450 pt]{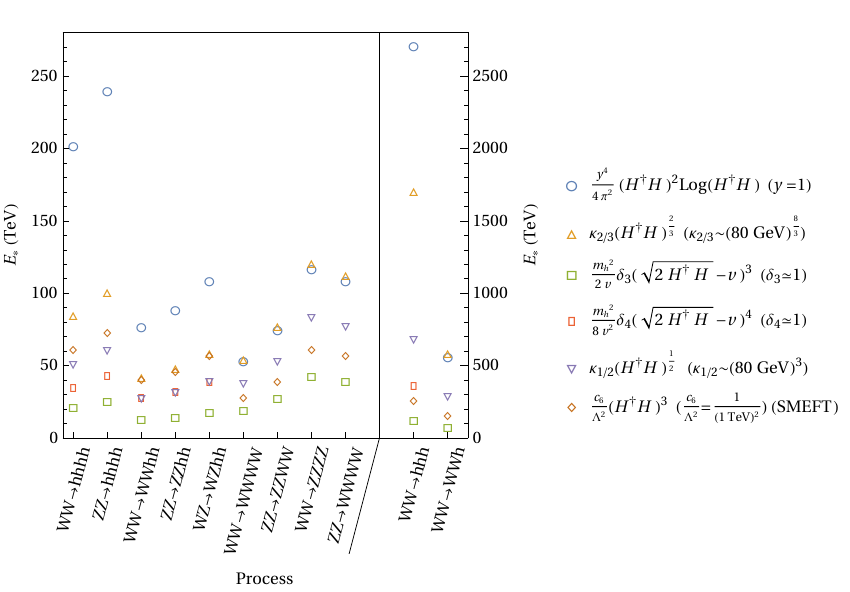}
\caption{Plot of $E_*$ values of multiple $2\to 4$ VBS processes, (left axis) as well as two chosen $2\to 3$ VBS processes (right axis). As discussed in Sec.~\ref{sec:Higgs-potential}, six different interactions are considered, including the quartic and cubic Higgs coupling modifications, as well as the dimension-six SMEFT operator. The quartic and cubic modifications (square and rectangle) are not turned on simultaneously, while the  other potentials (circle, diamond, triangles) predict both a cubic and quartic modification. If the Higgs cubic coupling is not modified (red rectangle), the $V_L V_L \to V_L^4$ and the $V_L V_L \to V_L V_L h$ processes have no $E_*$ value. The $E_*$ values in the 4-body final states are much lower than those in the 3-body final by approximately a factor of 10.}
\label{fig:res1}
\end{figure}

In this analysis, only the six-point constant contact term  is evaluated for the $2\to 4$ VBS, as any diagrams with propagators will necessarily be of a lower order in energy. From this contact term, we compute $\mathcal{M}$ to obtain s-wave component, $\hat{\cal M}$, and find $E_*$ by setting $|\hat{\mathcal{M}}|^2=1$. The energy growth comes from the phase space integration $\int \mathrm{dLIPS}_n$, giving an additional $E_{\rm cm}^4$ ($E_{\rm cm}^2$) term in $|\hat{\mathcal{M}}|^2$ in the $2\to 4$ ($2\to3$) case. In Fig.~\ref{fig:res1}, we plot $E_*$ values of the selected benchmark EFTs for the representative VBS processes. The tabulated results are found in Tables~\ref{table:res1} and \ref{table:res2} of App.~\ref{app:values24} for the $2\to4$ processes and in Table~\ref{table:res1old} of App.~\ref{app:values23} for the $2\to3$ processes.

The $2\to 4$ processes exhibit lower $E_*$ values compared to the $2\to 3$ processes, as the higher power of energy in $\hat{\cal M}$ compensates for the increased phase space suppression in $\mathrm{dLIPS}_4$, as discussed in Sec.~\ref{sec:PUV}. This trend is illustrated in Fig.~\ref{fig:res1}, where $E_{*, 2 \to 3} \sim 10 \cdot E_{*, 2 \to 4}$. Among the $2\to 4$ processes, the $V_L^4$ and $V_L V_L h h$ final states generally exhibit $E_*$ values that are equal to or smaller than that of the $h^4$ final state, except for the $W_L W_L \to Z_L^4$ and $Z_L Z_L \to W_L^4$ processes, which obtain larger symmetry factors compared to other final states involving $V_L$. If the cubic and quartic coupling deviations combine constructively, the $V_L V_L \to V_L V_L h h$ processes yield lower $E_*$ values than the $V_L V_L \to V_L^4$ processes. Conversely, when the deviations combine destructively, as observed in the logarithmic Higgs potential modification, the $W_L W_L \to W_L^4$ process results in the lowest $E_*$ value because $E_*$ values in $V_L^4$ final states depend solely on $\delta_3$. A similar cancellation also occurs for the $V_L V_L \to h^4$ processes which exhibit the highest $E_*$ values. 

As we show in Eq.~\eqref{eq:estarshifts}, the purely $V_L^4$ final states have no dependence on $\delta_4$, and consequently, these final states do not cause the tree-level unitarity violation if only the quartic coupling is modified, as seen in Fig.~\ref{fig:res1}. So, if $\delta_3$ is the sole quantity of interest, then the $V_L^4$ final states benefit from being independent of other modifications. This was also pointed out in Refs.~\cite{Chang:2019vez, Abu-Ajamieh:2020yqi}. However, if $\delta_4$ is of interest, it is probed by the $V_L V_L h h$ final states. The $V_L V_L hh$ final states also do not depend on $\delta_5$, so probing the $V_L^4$ and $V_L V_L h h$ final states would give enough information to disentangle both deviations of the Higgs potential. At the same time, the $h^4$ final state is sensitive to $\delta_5$. Therefore, probing the $h^4$ final state provides sufficient information to test for the quintic deviation. This is evident from Eq.~\eqref{eq:estarshifts}, as each set of processes exhibits distinct parametric dependencies on the Higgs self-couplings, offering compelling probes into the BSM theory. We discuss this detail for the cross sections in Sec.~\ref{sec:x-sec}.

While the obtained $E_*$ values for the $2\to4$ VBS is still out of reach at LHC, the significantly lower $E_*$ values compared to the $2\to3$ VBS motivates us to further investigate $2\to4$ VBS at the proposed future colliders, where the collision energy at the parton level could be as high as a few TeV. As we will discuss in Sec.~\ref{sec:x-sec}, the $2\to4$ VBS signal cross sections eventually exceed the $2\to3$ VBS signal cross sections at $\sqrt{\hat{s}} = {\cal O}(2$–$5)$\,TeV.  This advantage can be leveraged to enhance the sensitivity, in future colliders.

\section{Energy Growth for HEFTs with Derivative Modifications}\label{sec:HEFTderivative}

To explore more BSM scenarios, we consider sets of HEFT interactions involving derivatives. We select three benchmarks based on UV models which have a non-decoupling effect thus leading to the HEFT in the low energy. 
The two benchmark EFTs are from Loryon models introduced in Ref.~\cite{Banta:2021dek}, in particular a model with a singlet scalar and another one with a vector-like fermion. Additionally, based on the $(H^\dag H)^\frac{2}{3}$ potential briefly mentioned in Ref.~\cite{Falkowski:2019tft}, we also take the derivative terms from the UV model which would admit such a potential. The two models from Ref.~\cite{Banta:2021dek} also admit a potential term seen in Sec.~\ref{sec:Higgs-potential}, namely the $(H^\dag H)^2 \ln (H^\dag H)$ addition. 

Generically, we expect the matrix element, $\cal M$, from the derivative interactions to be of $\mathcal{O}(E_{\rm cm}^2)$. However, this is not always the case. In this section, we always consider the diagram which has the highest energy dependence. If $\mathcal{M} \propto E_{\rm cm}^2$ from the derivative terms, then the potential terms are neglected, since their matrix element contribution is always $E_{\rm cm}^0$. If the matrix element contribution from the derivative is of $E_{\rm cm}^0$ or less, we include the potential terms.

\subsection{EFTs from UV Models}
\label{subsec:uv}
The first UV model, briefly mentioned in Ref.~\cite{Falkowski:2019tft} is an additional $SU(2)$ doublet, $\Sigma$, with no bare mass,
\begin{equation}
\label{eq:fullsigma}
    \mathcal{L}_{\mathrm{UV}, \Sigma}= \mathcal{L}_{\mathrm{Higgs}}+|D \Sigma|^2+\kappa^2 (\Sigma^\dag H + \mbox{h.c.})-\lambda_\Sigma (\Sigma^\dag \Sigma)^2.
\end{equation}
Here, we choose the value of $\kappa$ and $\lambda_\Sigma$ based on bounds on $\kappa_3$, the Higgs cubic self-coupling, and making sure $0 < \lambda_\Sigma < 4 \pi$, while also constraining the model to be in the HEFT regime. This model can be integrated out using tree-level matching, giving the new EFT terms
\begin{equation}
    \mathcal{L}_{\mathrm{EFT}, \Sigma} = \mathcal{L}_{\mathrm{Higgs}}+\left( \frac{\kappa^2}{2 \lambda_\Sigma} \right)^\frac{2}{3} \frac{ \left|D H\right|^2 - \frac{8}{9}  \left(\partial \sqrt{H^\dag H}\right)^2}{\left(H^\dag H \right)^\frac{2}{3}}+\frac{3}{4} \left( \frac{\kappa^8}{\lambda_\Sigma} \right)^\frac{1}{3} \left( 2 H^\dag H \right)^\frac{2}{3}, 
\end{equation}
where we will be neglecting the $(H^\dag H)^\frac{2}{3}$ term, as the derivative term contribution to $\cal M$ always has a term dependent on $E_{\rm cm}^2$. As the potential term is being neglected, the only relevant parameter is $\left( {\kappa^2}/{ \lambda_\Sigma} \right)^\frac{1}{2} \equiv M = 45$ GeV. {The choice of the $\left(H^\dag H \right)^\frac{2}{3}$ potential in Sec.~\ref{subsec:HEFTpotential} corresponds to $\kappa_{2/3} \sim ( \kappa^8/ \lambda_\Sigma)^\frac{1}{3}$ $= (80$ GeV$)^{8/3}$. Assuming the same value of $M$, this corresponds to $\lambda_{\Sigma}\sim 5$, which remains within the perturbative range.}

Next, let us discuss two models from Ref.~\cite{Banta:2021dek} where the Higgs VEV significantly contributes to the mass of new fields. The first one introduces an additional singlet scalar, $S$,
\begin{equation}
    \mathcal{L}_{\mathrm{UV}, S}= \mathcal{L}_{\mathrm{Higgs}}-\frac{1}{2}S \left( \partial^2+m^2+\lambda H^\dag H \right) S. 
\end{equation}
The other model is a set of vector-like fermions, $\Psi$, with the Lagrangian 
\begin{equation}
 \mathcal{L}_{\mathrm{UV}, \Psi}= \mathcal{L}_{\mathrm{Higgs}}+ \bar \Psi \left( i \slashed{D} - m - \Xi \right) \Psi.
\end{equation}
where  $\Psi$ contains two Dirac fermions doublets, $\psi$ and $\chi$. $\psi$ transforms under $SU(2)_L$, and $\chi$ transforms under $SU(2)_R$. This means the mass term will be gauge-invariant. The $\Xi$ term mixes these doublets and adds a Higgs interaction, as 
\begin{equation}
    \Xi = \left( \begin{array}{c c} 0 & y \xi \\ y^* \xi^\dag & 0 \end{array}\right),
\end{equation}
and $\xi = \left( \begin{array}{c c} i \sigma_2 H^* & H \end{array}\right)$. This was done to enforce custodial symmetry. Based on the constraints in Ref.~\cite{Banta:2021dek}, turning off the bare mass for these theories are allowed. 

For the experimental bounds of these theories, see Ref.~\cite{Banta:2021dek} (Fig.~8, Fig.~10).  Note that the fermion model does not fit the bounds from $\kappa_\gamma$, the modification to the Higgs-photon-photon coupling. However, as the contribution to $\kappa_\gamma$ from these particles are negative, adding more particles can flip the sign of $\kappa_\gamma$ while maintaining the same magnitude.
To obtain the most prominent HEFT signatures, we set $m=0$ in both of these theories, and choose the $\lambda$ or $y$ benchmark based on what values they are allowed to be in the $m=0$ limit. 

Since in both models, the HEFT structure is not exhibited at tree level but appears at loop level, we use the effective potential formalism expanded to the two-derivative order to obtain this EFTs, as shown in App. D in \cite{Cohen:2020xca} or App. A in \cite{Banta:2021dek}. The HEFTs obtained for the scalar and fermion (in the massless limit), respectively, are 
\begin{align}
    \label{eq:scalar}
    \mathcal{L}_{\mathrm{EFT}, S} &= \mathcal{L}_{\mathrm{SM}}+\frac{\lambda}{96 \pi^2} \left(\partial \sqrt{H^\dag H}\right)^2+\frac{\lambda^2}{64 \pi^2} \left(H^\dag H \right)^2 \left(\ln \frac{\mu^2}{\lambda H^\dag H}+\frac{3}{2} \right),
    \\
     \mathcal{L}_{\mathrm{EFT}, \Psi} &= \mathcal{L}_{\mathrm{SM}}+\frac{y^2}{4 \pi^2} \left(
     |D H|^2\ln \frac{\mu^2}{y^2 H^\dag H} -\frac{2}{3}\left(\partial \sqrt{H^\dag H}\right)^2\right)- \frac{y^4}{4 \pi^2}(H^\dag H)^2 \left(\ln \frac{\mu^2}{y^2 H^\dag H}+\frac{3}{2} \right),
         \label{eq:fermion}
\end{align}
where $y$ is now real for simplicity.

As in the case for thase $SU(2)$ scalar doublet, the $\cal M$ derived from the fermion model always gives energy-growing behavior. Because of this, the potential term is neglected. However, for the singlet scalar, due to the vanishing bare mass, several processes have no energy growth in $\cal M$. For these processes, the potential terms are kept. Because of this, both cases seemingly have two relevant parameters. In the case of the scalar, the parameters are $\lambda$ and $\mu$, the renormalization scale. However, the $\mu$ dependence in the singlet scalar case only appears in the leading order for $2\to2$ processes, leading to only a single relevant parameter, $\lambda$. In the case of the fermion, the relevant parameters are $y$ and $\mu$. Here, we set $\mu=m_f \equiv y v/ \sqrt{2}$ as to minimize the logarithmic term.

\begin{figure}[!tbp]
\centering
\includegraphics[width=450 pt]{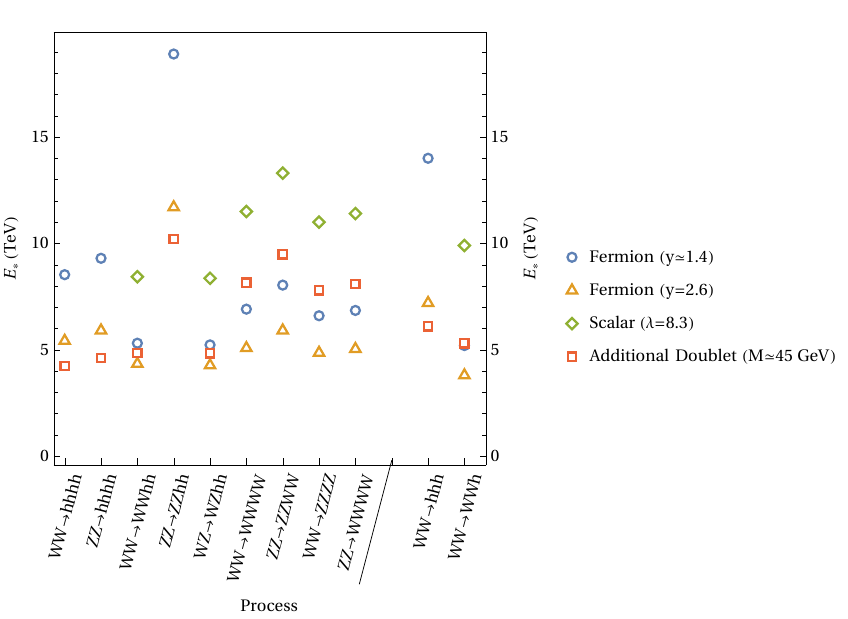}
\caption{
Plots of $E_*$ values of multiple $2\to 4$ processes, as well as two chosen $2\to 3$ processes. The HEFT interactions involve derivatives, as discussed in Sec.~\ref{sec:HEFTderivative}. The singlet scalar model is adopted from Ref.~\cite{Cohen:2021ucp}, but the parameter choice is different to set $m=0$ and to account the bounds in Ref.~\cite{Banta:2021dek}, as discussed in Secs.~\ref{subsec:uv} and~\ref{subsec:resder}. Here, $M \equiv \left( {\kappa^2}/{\lambda_\Sigma} \right)^\frac{1}{2}$, and $\mu=m_f \equiv y v / \sqrt{2}$, the mass of the fermion being integrated out. When going from the $2\to 3$ processes to $2\to4$ processes, there is little change in the order of $E_*$ values. The $E_*$ values from matrix elements of $\mathcal{O}(E_{\rm cm}^0)$, as in some scalar singlet cases, are an order of magnitude larger than those which are of $E_{\rm cm}^2$, and therefore have been excluded from the plot. For those values, see Table~\ref{table:res1d} in App.~\ref{app:values24}.}
\label{fig:res2}
\end{figure}

\subsection{Computing $\hat{\mathcal{M}}$}\label{subsec:mder}

In order to compute the matrix element, we apply the formalism developed in Ref.~\cite{Cohen:2021ucp}, as the new terms in the Lagrangian can be treated geometrically. Let us begin by writing the EFT's kinetic terms using the non-linear parametrization,
\begin{equation}
    \mathcal{L}_{\mathrm{kin}} = \frac{1}{2}K^2(h) \left(\partial h \right)^2+\frac{1}{2} v^2F^2(h) \left| \partial U \right|^2,
\end{equation}
where $H$ is expanded as $(v+h)U / \sqrt{2}$, and $U= \exp(i \sigma^a G^a/v)  \left(\begin{array}{c c} 0, & 1\end{array} \right)^T$, and $\sigma^a$ are the Pauli matrices. $U$ contains the NG bosons, and $h$ is the Higgs field. In the SM, $K(h)=1$, and $F(h)=1+\frac{h}{v}$. This means that $K(h)$ and $F(h)$ may make the theory non-canonically normalized if $K(0)$ or $F(0)$ are not $1$. As an example, let us see what $K(h)$ and $F(h)$ look like in the $SU(2)$ doublet addition,
\begin{equation}
\label{eq:doublet}
    \mathcal{L}_{\mathrm{kin},\Sigma} = \frac{1}{2}\left[1 + \frac{1}{9} \left( \frac{M}{v+h} \right)^\frac{4}{3} \right] \left( \partial h \right)^2 + \frac{1}{2}\left[1 + \left( \frac{M}{v+h} \right)^\frac{4}{3} \right] (v+h)^2 \left| \partial U \right|^2.
\end{equation}
Here, neither the Higgs nor the NG boson kinetic terms are canonical. However, the geometric method accounts for this, as all that is needed is dividing the final matrix element by $K(0)$ or $F(0)$ for each Higgs boson or NG boson in the process, respectively.  

From $K(h)$ and $F(h)$, the sectional curvatures are computed as 
\begin{equation}
\label{eq:kh}
    \mathcal{K}_h=-\frac{1}{K^2} \left( \frac{F''}{F} - \frac{K' F'}{K F}\right),
\end{equation}

\begin{equation}
\label{eq:kp}
    \mathcal{K}_\pi=-\frac{1}{v^2 F^2} \left( 1 - \frac{(v F')^2}{K^2}\right).
\end{equation}
Here, the prime denotes differentiation with $h$. Given these terms, the $V_L V_L \to h^4$ contact term is written as

\begin{equation}
\label{eq:nh}
     \mathcal{M}_{V_L V_L \to h^4} = -\left( s_{12} - \frac{2}{5} m_h^2\right) \left(\frac{\partial_h}{K(h)}\right)^{2} \mathcal{K}_h(h) \bigg|_{h=0},
\end{equation}
where $s_{12}= (p_1+p_2)^2=E_{\rm cm}^2$. Next, the $V_L V_L \to V_L V_L h h$ contact term is

\begin{align}
\label{eq:vv}
    \mathcal{M}_{V_L V_L \to V_L V_L h h} &= -\frac{1}{9}T_{ijkl} (s_{1234}-\frac{6}{5} m_h^2) \left(\frac{3 F'}{F K^2} \partial_h\right) \mathcal{K}_h(h) \bigg|_{h=0} + \left\{\frac{1}{6} T_{ijkl} s_{1234} \right.  \\ \nonumber
    &\left.-\frac{1}{2} \left[ \sigma_{ij} \sigma_{kl}(s_{12}+s_{34})+\sigma_{ik} \sigma_{jl}(s_{13}+s_{24})+\sigma_{il} \sigma_{jk}(s_{14}+s_{23}) \right] \right\} \left(\frac{\partial_h}{K}\right)^2\mathcal{K}_\pi(h) \bigg|_{h=0},
\end{align}
where $T_{i j k l}$ and $\sigma_{i j}$ are defined as
\begin{align}
\label{eq:sigma}
T_{i j k l} = \left( \sigma_{ij} \sigma_{kl} + \sigma_{ik} \sigma_{jl} + \sigma_{il} \sigma_{jk}\right), \quad  
    \sigma = \left(\begin{array}{c  c  c} 0 & 1 & 0 \\ 1 & 0 & 0 \\ 0 & 0 & 1 \end{array} \right),
\end{align}
and $s_{ij...k}=(p_i+p_j+\cdots p_k)^2$, where all momenta point inwards.  

For the $V_LV_L \to V_L^4 $ case, the more generic formula in terms of the Riemann tensor was applied from Eq.~(2.22) of Ref.~\cite{Cohen:2021ucp}, 

\begin{equation}
\label{eq:6vgen}
    F(0)^6 \mathcal{M}_{a_1 a_2 a_3 a_4 a_5 a_6} = \frac{3}{5}\sum_{1\leq i < j \leq 6} s_{ij} R_{a_i(a_1 a_2|a_j|;a_3\dots \hat{a}_i \hat{a}_j \dots a_6)} \bigg|_{h=0} + \mathcal{O}(R^2).
\end{equation}
Here, $R_{ijkl}$ is the Riemann tensor, the semi-colon denotes covariant differentiation, and the $a_i$ would be replaced by whichever NG boson is of interest in the $(\pi_1, \pi_2, \pi_3)$ basis. The indices inside the parentheses are symmetrized, and hatted indices and the index within the vertical bars (|···|) must be excluded from the symmetrization. The $\mathcal O(R^2)$ terms are being excluded, as Ref.~\cite{Cohen:2021ucp} applies a formalism which truncates the expression at $\mathcal{O}(R^2)$, making Eq.~\eqref{eq:6vgen} valid for the lowest order coupling of the EFT operator. To check how well this approximation holds, let us apply it in the case of the loop-level singlet loryon by looking at ${\cal K}_\pi$, as $R_{ijkl} = (g_{ik}g_{jl}-g_{il}g_{jk}) {\cal K}_\pi$, and $g_{ij} \propto F(h)=1+\frac{h}{v}$ (here, $i,j,k,l$ indicate NG boson indices). If $m=0$,
\begin{equation}
    \mathcal{K}_\pi= \frac{\lambda}{96 \pi^2}  \frac{1}{(1+\frac{\lambda}{96 \pi^2})(v+h)}.
\end{equation}
Any $\mathcal O(R^2)$ term will necessarily have higher powers of the factor ${\lambda}/{96 \pi^2}$, so Eq.~\eqref{eq:6vgen} gives the leading order contribution of $\mathcal{O}\left( {\lambda}/{96 \pi^2} \right)$, assuming $\lambda \ll 96 \pi^2$. In general, as long as the new coupling, including numerical factors, is much smaller than $\mathcal O(1)$, the $\mathcal{O}(R^2)$ approximation will hold.  As a consequence of keeping the computation $\mathcal O(R^2)$, the covariant derivatives are assumed to be symmetric. As the commutator of the covariant derivatives applied to a generic tensor, $T$, is of $\mathcal{O}(RT)$, the commutator of the covariant derivatives on the Riemann tensor is $\mathcal{O}(R^2)$.

The result in Eq.~\eqref{eq:6vgen} is still in the $(\pi_1, \pi_2, \pi_3)$ basis, so to go to the $(W^+_L, W^-_L, Z_L)$ basis, we must apply the transformation
\begin{equation}
\label{eq:sigmat}
    \tilde \sigma = \left(\begin{array}{c  c  c} \frac{1}{\sqrt{2}} & \frac{-i}{\sqrt{2}} & 0 \\ \frac{1}{\sqrt{2}} & \frac{i}{\sqrt{2}} & 0 \\ 0 & 0 & 1 \end{array} \right),
\end{equation}
on every $\pi$ index in Eq.~\eqref{eq:6vgen}. This will give the final result for the $V_L V_L \to V_L^4$ matrix element. 

While a closed form solution is not stated, $\mathcal{M}_{V_L V_L \to V_L^4} \propto  \mathcal{K}_\pi(h) /[v F(h)]^2 \big|_{h=0}$. This means there is no dependence on $\mathcal{K}_h$, unlike in the $V_L V_L \to V_L V_L h h $ case. This leads to slightly higher $E_*$ values in the vector boson-only final states.

Eqs.~(\ref{eq:nh}, \ref{eq:vv}, \ref{eq:6vgen}) make the $E_{\rm cm}$, momentum and angular dependences of the processes manifest. This is most apparent in the case of the NG boson scattering examples. It is interesting to note that the $h^4$ final state has no angular or momentum dependence in the contact term. This means the matrix element remains factorizable from the phase space integral, as there is no integration over $E_{\rm cm}$. In the other cases, however, the phase space integral, $\int \mathrm{dLIPS}_4 \mathcal{M}$, can \textit{not} be factorized, which leads to a non-trivial integral, evaluated using a Monte Carlo integration. If $\cal M$ depends on the phase space variables, the integral $\int \mathrm{dLIPS}_4 \cal M$ is smaller than the case where $\cal M$ is constant. Hence, the final states involving vector bosons have an extra suppression factor in $\hat{\cal M}$.

 For diagrams with internal propagators, the phase space integral provides numerical suppression. Although these diagrams exhibit energy-growing behavior beyond the contact terms, this suppression allows us to neglect them when determining $E_*$ safely. This is also shown in Ref.~\cite{Delgado:2023ynh}, where the propagator terms in the $V_L V_L \to h^4$ corresponded to an $\mathcal{O}(10\%)$ shift. As in the case of derivatives, $E_*\propto | \mathcal{M}|^{-{1}/{4}}$, this will correspond to a deviation of $\mathcal{O}(2\%)$ in $E_*$.  

\subsection{Results}
\label{subsec:resder}

The $E_*$ values from the resulting contact terms are given in Fig.~\ref{fig:res2}, and the tabulated values with extra processes are given in Tables~\ref{table:res1d} and \ref{table:res2d}. {For completeness, in Table~\ref{table:res22}, we also present $E_*$ values for the $2\to 2$ processes. We justify the approximation where the contributions from the non-analytic potential are dropped because including those modifies the $E_*$ values at most 1\%.}  Relevant $2 \to 3$ VBS processes are also included for comparison with the $2 \to 4$ processes, as shown in Table~\ref{table:res23}. Among the $2\to 4$ processes, most of the $E_*$ values are similar, being of ${\cal O}(5$-$10$)\,TeV. The process with the lowest $E_*$ value depends on the chosen EFT. For example, in the fermion (triangle and circle) and singlet scalar (diamond) cases shown in Fig.~\ref{fig:res2}, the $W_L Z_L \to W_L Z_L h h$ process has the lowest $E_*$ value, while for the doublet scalar (square), the $W_L W_L \to h^4$ process has the lowest $E_*$ value. An interesting aspect of the $V_L V_L \to V_L V_L h h$ processes is that they are sensitive to both ${\cal K}_h$ and ${\cal K}_\pi$ as shown in Eq.~\eqref{eq:vv}, unlike the $V_L^4$ and $h^4$ final states, which are sensitive only to ${\cal K}_\pi$ and ${\cal K}_h$, respectively.

Compared to the $2\to3$ case, the $h^4$ final state has a lower $E_*$ value than the $hhh$ final state. However, the $E_*$ value under HEFT derivative modifications does not drop as drastically as in the modified Higgs potential, which is seen when comparing Tables~\ref{table:res1d} and~\ref{table:res23}. This is because the lower multiplicity final states still have energy growth in the matrix element for the derivative HEFT operators. Even the $2\to2$ processes have terms in $\cal M$ proportional to $E_{\rm cm}^2$. However, despite the smaller drop in $E_*$, in the case of $V_L V_L \to h^n$, there is a clear trend of smaller $E_*$ values when moving to the $n = 4$ case, as seen by comparing the $hhh$ and $h^4$ final states in Fig.~\ref{fig:res2} This trend is not followed by the final processes involving vector bosons, as for those processes, the $E_*$ values actually increase when going from the 3-body final states to the 4-body final states. Hence, for derivative HEFT operators, going to higher multiplicity final states involving vector bosons does not have a benefit compared to the potential modifications. Nevertheless, higher multiplicities using the $h^4$ final state remain appealing.

Note that some of the $E_*$ values for the singlet scalar are not shown in Fig.~\ref{fig:res2}, namely for the $V_L V_L \to h^4$ and $Z_L Z_L \to Z_L Z_L h h$ processes, because their matrix elements are proportional to $E_{\rm cm}^0$, resulting in much larger $E_*$ values. In the case of $V_L V_L \to h^4$, this occurs due to the parameter choices for the UV model, specifically setting $m = 0$. As $\mathcal{K}_h \propto m^2$ in the case of the singlet scalar, turning off the bare mass causes the term proportional to $E_{\rm cm}^2$ in the matrix element to vanish. This is the same reason why $V_L V_L \to hhh$ has no $E_*$ value in the plot for the singlet scalar. The same model was studied in Ref.~\cite{Cohen:2021ucp} in the context of $V_L V_L \to h^n$ scattering, but with the bare mass kept on, leading to $\cal M$ having terms of ${\cal O}(E_{\rm cm}^2)$. While the $Z_L Z_L \to Z_L Z_L h h$ scattering has a $\mathcal{K}_\pi$ dependence, the momentum dependence associated with $\mathcal{K}_\pi$ in Eq.~\eqref{eq:vv} results in a term proportional to $m_Z^2$. For this reason, even in the other EFTs, the $E_*$ values of the $Z_L Z_L \to Z_L Z_L h h$ process are much higher than the other $V_L V_L h h$ final states in Fig.~\ref{fig:res2}, as only $\mathcal{K}_h$ contributes a term proportional to $E_{\rm cm}^2$ to the matrix element.

The fact that some $2\to4$ processes have $E_*$ values of the same order as the $2\to3$ processes under HEFT derivative modifications leads to consequences in the signal cross section. As in Eq.~\eqref{eq:x-sec_scaling}, the cross section can be schematically written as $E_{\rm cm}/E_*$ raised to a power. We also show that higher multiplicity leads to a higher power of $E_{\rm cm}/E_*$ in the cross section for Higgs potential modifications. Generically, this holds for the derivative modifications as well. However, if the $E_*$ values for $2 \to 3$  and $2 \to 4$ processes are similar, the $2 \to 4$ process will experience a higher power suppression of $E_{\rm cm}/E_*$, as $E_{\rm cm}$ must remain below $E_*$. For this reason, the signal cross section in the final states involving vector bosons will not be larger in the $2\to 4$ case compared to the $2\to 3$ case. In contrast, since the $h^4$ final state still has a smaller $E_*$ value than the $hhh$ final state, this conclusion does not necessarily apply to the Higgs-only final states. The cross section will be examined in greater detail in the following section, Sec.~\ref{sec:x-sec}.

\section{Cross section for a Matrix Element Compared to $E_*$}\label{sec:x-sec}

In this section, we apply our analysis of the $E_*$ values from Secs.~\ref{sec:Higgs-potential} and~\ref{sec:HEFTderivative} to see their effects on the BSM signal cross sections. The cross sections of higher multiplicity processes could have stronger signals beyond certain parton energy thresholds. We will show that the 4-body final states expect more signal yields than the 3-body ones at the parton energies of a few TeV if solely the Higgs potential is modified. This situation could be realized in high-energy future colliders. On the other hand, in the presence of HEFT derivative operators, the $2\to4$ VBS tends to have a lower cross section than the $2\to3$ VBS within the valid regime of EFT.

\subsection{Constant Matrix Element}

From our criteria of perturbative unitarity violation, $E_*$ is defined as the $E_{\rm cm}$ value when Eq.~\eqref{eq:mhatcondition} applies, i.e. for a constant matrix element, setting Eq.~\eqref{eq:Mavg} to unity and setting $m=2$ gives,
\begin{equation}
\label{eq:emhat}
    |\hat{\mathcal{M}}|^2 = \frac{1}{ S_{\mathrm{in},2} \cdot S_{ \mathrm{out},n}} \frac{1}{8 \pi} \int \mathrm{dLIPS}_n |\mathcal{M}|^2 \bigg.|_{E_{\rm cm}=E_*} =1.
\end{equation}
Next, from the cross section formula in the massless limit, for a constant matrix element (as in Sec.\ref{sec:Higgs-potential}), one writes
\begin{equation}
\label{eq:exsec}
    \sigma = \frac{1}{S_{ \mathrm{out},n}} \frac{1}{2 E_{\rm cm}^2} \int \mathrm{dLIPS}_n |\mathcal{M}|^2,
\end{equation}
Then, dividing Eq.~\eqref{eq:exsec} by Eq.~\eqref{eq:emhat} gives

\begin{equation}
\label{eq:exsec2}
    \sigma = 4 \pi S_{ \mathrm{in},2} \frac{E_{\rm cm}^{2(n-3)}}{E_*^{2(n-2)}},
\end{equation}
As the same initial state of $W_L W_L$ is used for all processes in this section, $S_{ \mathrm{in},2}$ is the same in all cases. The cross section for the 3-body and 4-body final states for a constant matrix element can be obtained by setting $n=3,4$, so

\begin{equation}
    \sigma_3= 4 \pi S_{\mathrm{in},2} \frac{1}{E_{*, 2 \to 3}^2},
\end{equation}
\begin{equation}
    \sigma_4= 4 \pi S_{\mathrm{in},2} \frac{E_{\rm cm}^2}{E_{*, 2 \to 4}^4}.
\end{equation}
Here, $E_{*, 2 \to n}$, $n=3,4$ are the respective $E_*$ values. These two expressions can now be equated to find $E_\sigma$, the approximate (parton) energy at which $\sigma_4$ catches up to and becomes larger than $\sigma_3$. This gives
\begin{equation}
\label{eq:ec}
    E_\sigma= \frac{E_{*, 2 \to 4}^2}{E_{*, 2 \to 3}}.
\end{equation}
From the results of Fig.~\ref{fig:res1} and using the $W_L W_L \to h^n$ processes with a cubic coupling deviation as an example, Eq.~\eqref{eq:ec} ends up giving $E_\sigma= 3.6$ TeV, which may be reachable in future colliders.  If there is no quintic term modification, this value is \textit{independent} of the quartic and cubic deviations. This is true even if only a quartic shift was assumed. This can be explained by the fact that $E_{*, 2 \to 3}$ has a factor of $|\delta_3- \frac{1}{3}\delta_4|$ in the denominator, and $E_{*, 2 \to 4}$ has a $|\delta_3- \frac{1}{3}\delta_4|^{{1}/{2}}$ if $\delta_5=0$. Hence, $E_{*}^2$ for the $W_L W_L \to h^4$ process has the same denominator as the $E_*$ for the $W_L W_L \to h h h$ process. This means ${E_{*, 2 \to 4}^2}/{E_{*, 2 \to 3}}$ will not have any dependence on $\delta_i$. {This can be seen explicitly if we write the cross sections in terms of $\delta_i$,
\begin{align}
\label{eq:hhhxsec}
    \sigma_{W_LW_L\to hhh} &= 0.34  \left(\delta_3-\frac{1}{3} \delta_4\right)^2  \mbox{ pb}, \\ \label{eq:4hxsec}
    \sigma_{W_LW_L\to h^4} &= 0.025 \left(\frac{E_{\rm cm}}{\mbox{TeV}}\right)^2  \left(\delta_3-\frac{1}{3} \delta_4+ \frac{10}{3} \frac{v^2}{m_h^2} \delta_5 \right)^2 \mbox{ pb}.
\end{align}
Hence, if $\delta_5=0$, the two cross sections have the same dependence on the modifications.} 

The independence of $E_\sigma$ on the coupling deviation also holds when comparing the $V_L^4$ final states to the $V_L V_L h$ final states, as their $E_*$ value denominators are $|\delta_3|^{{1}/{2}}$ and $|\delta_3|$ respectively. {The cross sections are written as
\begin{align}
\label{eq:wwhxsec}
    \sigma_{W_LW_L\to W_L W_L h} &= 0.97 \cdot \delta_3^2  \mbox{ pb}, \\ \label{eq:4wxsec}
    \sigma_{W_LW_L\to W_L^4} &= 0.038\left(\frac{E_{\rm cm}}{\mbox{TeV}}\right)^2   \delta_3^2 \mbox{ pb}.
\end{align}
}

The independence of couplings is not true when comparing the $V_L V_L h h$ final states to any other $2\to 3$ processes unless $\delta_4$ is turned off, as its $E_*$ value has a unique denominator{. This is seen in its explicit cross section when compared to Eqs.~(\ref{eq:hhhxsec}, \ref{eq:wwhxsec}):
\begin{equation}
\label{eq:wwhhxsec}
    \sigma_{W_L W_L \to W_L W_L hh} =0.21 \left(\frac{E_{\rm cm}}{\mbox{TeV}}\right)^2  \left( \delta_3- \frac{1}{5} \delta_4 \right)^2 \mbox{ pb}.
\end{equation}
}For other general parametric dependences on $\delta_i$ in $E_*$, see Eq.~\eqref{eq:estarshifts} and Tables~\Ref{table:res1} and~\Ref{table:res1old}. 

Note that the $E_\sigma$ value is before the background is taken into account. If the background for the $V_L V_L \to h^4$ processes are smaller than the $V_L V_L \to hhh$ processes, the signal will give a large significance value even before reaching $E_\sigma$. While the precise shape of the background is unknown, having a higher multiplicity in the final state leads to more phase space factors of the $2\to4$ process. This could make its background cross section smaller than the $2\to3$ process. Hence, even with similar signal sizes, the $2\to4$ process can be expected to have higher significance values before reaching $E_\sigma$. For purely scalar scattering states, it is also expected that the cross section is decreasing with energy, so going to higher energies may lead to even smaller backgrounds for the $h^4$ case compared to the $hhh$ case, leading to an even stronger significance. 

\begin{figure}[!tbp]
\centering
 {\includegraphics[width=0.5\textwidth]{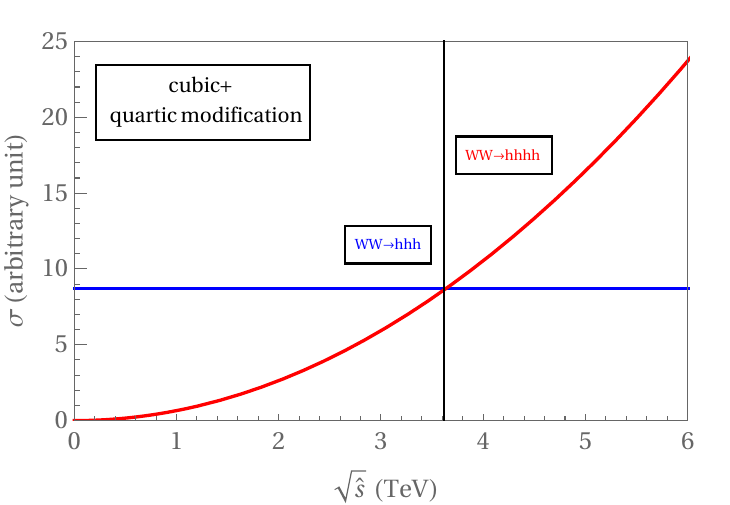}}~
 {\includegraphics[width=0.5\textwidth]{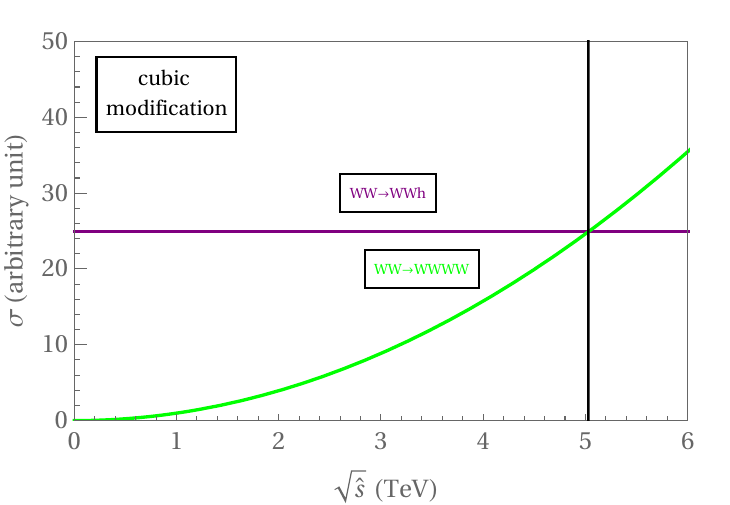}}
\caption{Signal cross section of the BSM part of $W_L W_L \to hhh$  and  $W_L W_L \to h^4$ processes with Higgs coupling modifications of $\delta_{3,4}\neq 0$ and $\delta_{5}=0$ in an arbitrary unit  (left), and signal cross section of the BSM part of  $W_L W_L \to W_LW_Lh$ and $W_L W_L \to W_L W_L W_L W_L$ processes  with $\delta_{3}\neq 0$ in an arbitrary unit as discussed in Sec.~\ref{sec:Higgs-potential}(right). The horizontal axis is the center of mass energy at parton level, $\sqrt{\hat s}$. The vertical lines indicates $E_\sigma$ above which the $2\to 4$ VBS cross section exceeds the one of the $2\to 3$ VBS. In both plots, $E_\sigma$  is independent of specific values of $\delta_{3,4}$, {as shown using Eq.~\eqref{eq:ec} and in Eqs.~(\ref{eq:hhhxsec},~\ref{eq:4hxsec},~\ref{eq:wwhxsec},~\ref{eq:4wxsec}).}} 
\label{fig:potentialh}
\end{figure}

\begin{figure}[!tbp]
\centering
 {\includegraphics[width=0.5\textwidth]{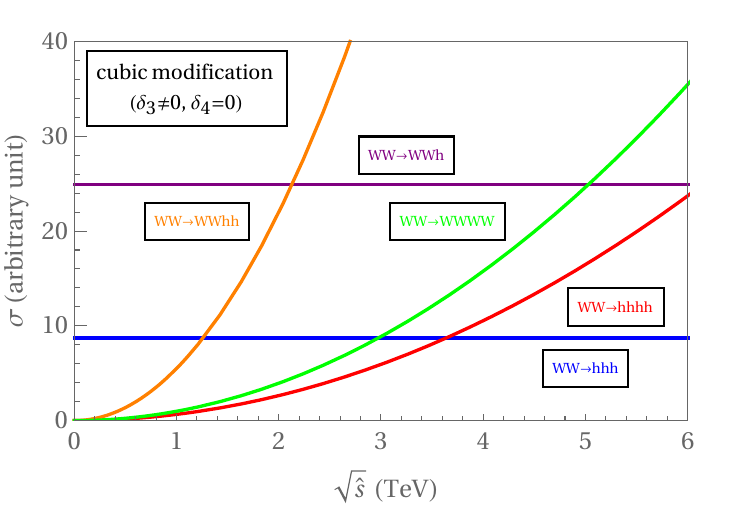}}~
 {\includegraphics[width=0.5\textwidth]{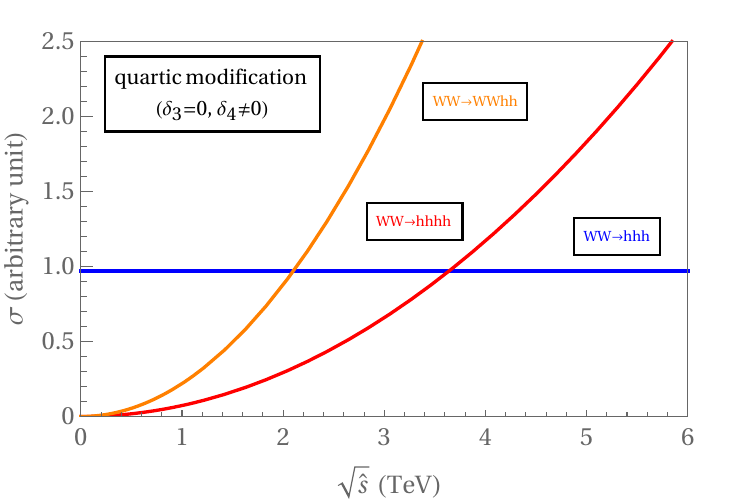}}
\caption{Signal cross section in an arbitrary unit with only a cubic coupling modification $(\delta_3\neq 0, \delta_{4}=0)$  (left), and signal cross section in an arbitrary unit assuming only a quartic coupling modification $(\delta_4\neq 0, \delta_{3}=0)$ (right). The quintic coupling is absent, $\delta_5=0$. Not all processes are sensitive to the quartic coupling deviation, so only those which have a quartic deviation dependence are shown in the right plot. As is the case with Fig.~\ref{fig:potentialh}, the value of $E_\sigma$ is independent of the couplings when comparing $h^4$ with $hhh$ and $W_L^4$ with $W_L W_L h$, {as discussed under Eq.~\eqref{eq:ec}, and in Eqs.~(\ref{eq:hhhxsec},~\ref{eq:4hxsec},~\ref{eq:wwhxsec},~\ref{eq:4wxsec}).} Some $2\to 4$ VBS can have more signal events than $2\to3$ VBS as low as $\sqrt{\hat s}\simeq 2$~TeV. } 
\label{fig:potentialall}
\end{figure}

For the exact energy-growing behavior of the BSM part of the $2\to 4$ cross section, see Fig.~\ref{fig:potentialh}. In addition to comparing the Higgs-only final states in the HEFT scenario, Fig.~\ref{fig:potentialh} right is a comparison of the $W_L W_L \to W_L^4$ cross section with the $W_L W_L \to W_L W_L h$ cross section. These final states have similar signal behavior as the $h^4$ and $hhh$ final states, albeit $E_\sigma$ is slightly higher. As seen from Eqs.~(\ref{eq:estarshifts}, \ref{eq:ec}), there is a cancellation between the $\delta_3$ dependences of the $E_*$ values, which are independent of turning on $\delta_{n \geq 4}$. Fig.~\ref{fig:potentialall} compares all processes with each other, given either only cubic modification (left) or only quartic modification (right). In Fig.~\ref{fig:potentialall} right, only the processes sensitive to $\delta_4$ are plotted. 

\begin{figure}[!tbp]
\centering
 {\includegraphics[width=0.5\textwidth]{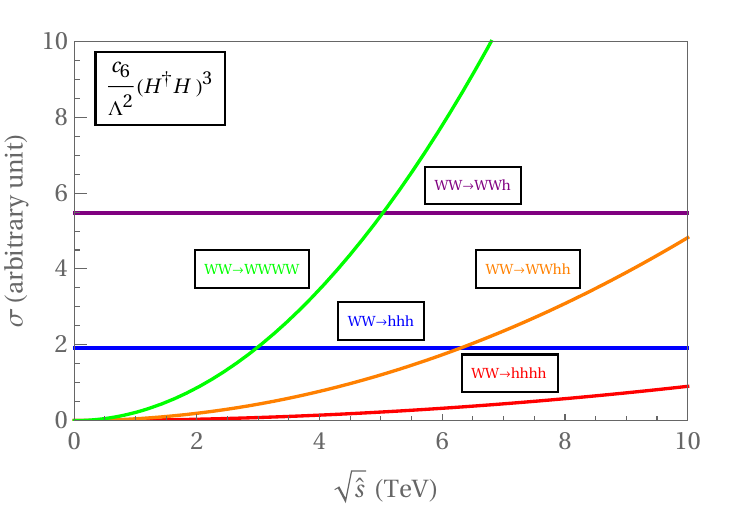}}~
\caption{Similar plot as Fig.~\ref{fig:potentialall}, but now with the coupling modifications from the dimension-six SMEFT operator, which leads to the correlations shown in Eq.~\eqref{eq:smeftcorrelation}. From cancellations caused by the new deviations, the $h^4$ and the $W_L W_L h h$ final states have much larger $E_\sigma$ values compared to the $\delta_3 \neq 0$, $\delta_{4,5}=0$ case. However, the $W_L W_L \to W_L^4$ process has the same $E_\sigma$ as before. The value of $(c_6 / \Lambda^2)$ does not affect the $E_\sigma$ value.} 
\label{fig:potentialsmeft}
\end{figure}

In Fig.~\ref{fig:potentialsmeft}, The dimension-six SMEFT caveats from Eq.~\eqref{eq:smeftcorrelation} are applied. Now, from cancellations between $\delta_3$, $\delta_4$, and $\delta_5$, some of the processes have weaker energy growth, such as the $h^4$ final state, which had its $E_\sigma$ value changed by a factor of 4 compared to the $\delta_3 \neq 0$, $\delta_{4,5}=0$ case, and the $W_L W_L \to W_L W_L hh$ process, which saw it changed by a factor of 5. The $W_L W_L \to W_L^4$ process had no such change, as the only important modification is $\delta_3$, so including and excluding any correlations causes no change in $E_\sigma$ when comparing the $W_L W_L h$ and $W_L^4$ final states. The value of $c_6 / \Lambda^2$ does not affect $E_\sigma$.

Comparing the plots of the SMEFT case to the two HEFT cases shows how extra probes which can access the quartic can better disentangle deviations of the Higgs potential. Even with similar values of $\delta_3$, the cross sections of a process can look different if it is sensitive to the other Higgs potential modifications. With the dimension-six SMEFT operator, the presence of $\delta_3$ necessitates nonzero values for all $\delta_i$ with $i = 4$ to $6$, inherently making predictions for higher-order deviations. The resulting changes in these observables provide deeper insight into the nature of the underlying theory. For example, if the cross section of the $W_L W_L \to W_L W_L hh$ process looks more similar to Fig.~\ref{fig:potentialall} rather than Fig.~\ref{fig:potentialsmeft}, the SMEFT can not hold, as turning off the higher order self-couplings breaks the expected relations. A similar statement may be said for dimension-eight operator, as it will induce $\delta_i$ for $i=$ 5 to 8, though only the processes with a quintic modification dependence can probe this, such as the $h^4$ final state.  

\subsection{Derivative Modifications}
\label{subsec:derxsec}

\begin{figure}[!tbp]
\centering
{\includegraphics[width=0.5\textwidth]{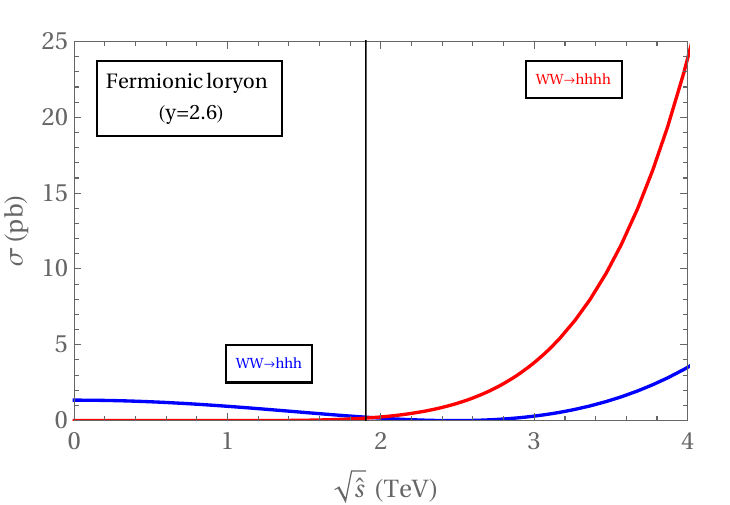}}~
 {\includegraphics[width=0.5\textwidth]{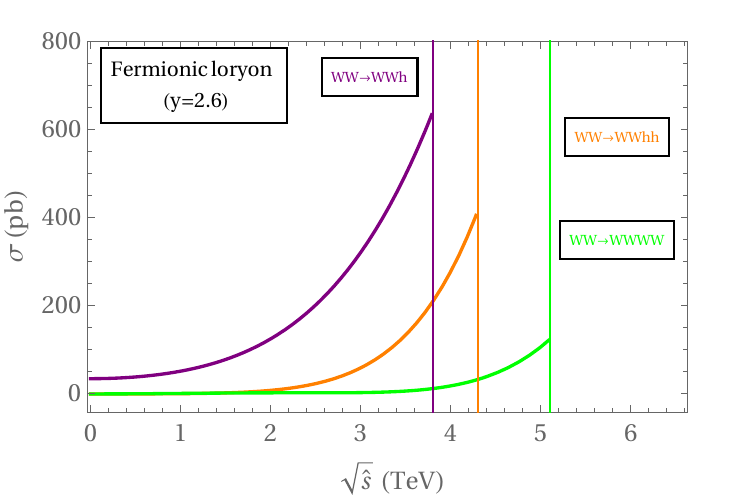}}
\caption{Signal cross sections in picobarns for the HEFT with derivatives, as defined by Eq.~\ref{eq:fermion}, with $y=2.6$. On the left, the $W_L W_L \to h h h$ and the $W_L W_L \to  h^4$ processes are shown. {This corresponds to Eqs.~(\ref{eq:3hxs}, \ref{eq:4hxs})}. On the right, two $2\to4$ processes, $W_L W_L \to W_L W_L h h$ and $W_L^4$, are shown, along with the $W_L W_L \to W_L W_L h$ process. {This corresponds to Eqs.~(\ref{eq:wwhderxsec}, \ref{eq:4wxs}, \ref{eq:wwhhxs})}. On the right plot, the vertical lines indicate the process's $E_*$ value, after which the EFT description breaks down. Unlike the Higgs-only final state case, the $W_L W_L \to W_L W_L h$ cross section remains larger than the cross sections of $W_L W_L \to W_L W_L hh, W_L^4$ in the regime where the EFT is valid.
} 
\label{fig:potentialdw}
\end{figure}

In general, derivative operators can induce non-trivial momentum dependences in the matrix element, so $\int \mathrm{dLIPS}_n \mathcal{M}$ does not usually have a closed-form expression. However, the simplest case, that of $V_L V_L \to h^4$ has an analytic expression, as $\cal M$ is factorizable from $\mathrm{dLIPS}_n$. {To begin, let us neglect the potential term of the matrix element, and apply} Eq.~\eqref{eq:nh},
\begin{equation}
\label{eq:prenh}
     \mathcal{M}_{V_L V_L \to h^4} \propto \left( s_{12} - \frac{2 m_h^2}{5} \right) \simeq E^2.
\end{equation}
This means $\mathcal{M}$ is constant with respect to $\mathrm{dLIPS}_n$, and a similar equation to Eq.~\eqref{eq:emhat} holds,

\begin{equation}
\label{eq:emhatder}
    |\hat{\mathcal{M}}|^2 = \frac{1}{ S_{\mathrm{in},2} \cdot S_{ \mathrm{out},n}} \frac{1}{8 \pi} \int \mathrm{dLIPS}_n |\mathcal{M}|^2 \bigg.|_{E_{\rm cm}=E_*} = \frac{1}{ S_{\mathrm{in},2} \cdot S_{ \mathrm{out},n}} \frac{1}{8 \pi} E_*^{2n} M_n= 1,
\end{equation}
where $M_n$ encodes the proportionality term of $|\mathcal{M}|^2$ as well as the part of the phase space integral independent of energy. Similarly, the cross section {from the derivative terms, $\sigma_D$,} is written as 
\begin{equation}
\label{eq:exsecder}
    {\sigma_D} = \frac{1}{S_{\mathrm{out},n}} \frac{E_{\rm cm}^{2(n-1)}}{2} M_n,
\end{equation}
and dividing Eq.~\eqref{eq:exsecder} by Eq.~\eqref{eq:emhatder},
\begin{equation}
\label{eq:exsec2der}
    {\sigma_D} = 4 \pi S_{\mathrm{in},2} \frac{E_{\rm cm}^{2(n-1)}}{E_*^{2n}}.
\end{equation}
{As the $E_\sigma$ value is at low energies, around $\cal O$(1) TeV for the $W_L W_L \to h^n$ processes, we include the potential and interference contributions in addition in addition to $\sigma_D$. Although the leading contribution to the $W_L W_L \to h^n$ cross section is proportional to $y^8$, we retain all higher powers of $y$ in the numerical evaluation. At $y=2.6$, the cross sections are
 \begin{align}
 \label{eq:3hxs}
\sigma_{W_LW_L\to hhh} &= \left[37-6.0  \left(\frac{E_{\rm cm}}{\mbox{TeV}}\right)^2 \right]^2   \mbox{ fb}, \\ \label{eq:4hxs}
    \sigma_{W_LW_L\to h^4} &= \left[ 2.5 \left(\frac{E_{\rm cm}}{\mbox{TeV}}\right)-2.6 \left(\frac{E_{\rm cm}}{\mbox{TeV}}\right)^3 \right]^2 \mbox{ fb}.
\end{align}  
Setting the expressions to be equal gives $E_\sigma = 1.9$~TeV. Hence, for Higgs-only final states, the $2\to4$ process still leads to a larger BSM cross section. The cross sections from Eqs.~(\ref{eq:3hxs}, \ref{eq:4hxs}) can be seen in Fig.~\ref{fig:potentialdw} on the left.}

 However, in the case of the final states involving vector bosons, the situation changes. As $\cal M$ is not factorizable from $\mathrm{dLIPS}_n$, a Monte Carlo integration is used to evaluate the phase space integrals. {We evaluate the cross sections for the $W_L W_L \to W_L W_Lh$, $W_L W_L \to W_L^4$, and $W_L W_L \to W_L W_L hh$ processes for $y=2.6$,  
\begin{align}
\label{eq:wwhderxsec}
\sigma_{W_LW_L\to W_L W_Lh} &= \left[35+15  \left(\frac{E_{\rm cm}}{\mbox{TeV}}\right)^2+1.9  \left(\frac{E_{\rm cm}}{\mbox{TeV}}\right)^4 \right]   \mbox{ pb}, \\ \label{eq:4wxs}
    \sigma_{W_LW_L\to W_L^4} &=  \left[ 1400 \left(\frac{E_{\rm cm}}{\mbox{TeV}}\right)^2 - 250 \left(\frac{E_{\rm cm}}{\mbox{TeV}}\right)^4+14  \left(\frac{E_{\rm cm}}{\mbox{TeV}}\right)^6 \right] \mbox{ fb}, \\ \label{eq:wwhhxs}
     \sigma_{W_LW_L\to W_L W_L hh} &=  \left[ 310 \left(\frac{E_{\rm cm}}{\mbox{TeV}}\right)^2 + 220 \left(\frac{E_{\rm cm}}{\mbox{TeV}}\right)^4+ 53 \left(\frac{E_{\rm cm}}{\mbox{TeV}}\right)^6 \right] \mbox{ fb}.
\end{align}
where the potential contributions are also kept to present the low-energy regime. 
The dependence on $y$ is more involved as the potential terms and $\mathcal{K}_h$ are proportional to $y^4$ while $\cal K_\pi$ is proportional to $y^2$. The matrix elements for these processes involve all three terms, as can be seen in Eqs.~(\ref{eq:log}, \ref{eq:vv}, \ref{eq:6vgen}). }
In Fig.~\ref{fig:potentialdw} right, we show that these 4-body final states always have a smaller cross section than the 3-body final state within the valid energy, $\sqrt{\hat{s}}<E_{*,2\to4}$. Hence, the $E_\sigma$ value is undefined when comparing the $W_L W_L hh$ and $W_LW_Lh$ final states, as well as the $W_L^4$ and $W_LW_Lh$ final states. As a consequence of this behavior, the $W_L W_L \to W_L W_L h$ process is preferred over the $W_L W_L \to W_L W_L h h$ and $W_L^4$ processes. 
 For some UV models, if final states involving vector bosons are of interest, even the $2\to 2$ processes can be a good probe into new physics involving derivative modifications.

\section{Discussion and Conclusion}

Identifying sensitive and independent processes to probe the Higgs cubic and quartic couplings is crucial for advancing our understanding of the underlying new physics. While di-Higgs production remains the most promising probe of the cubic coupling, achieving higher precision requires combining it with other complementary measurements. For the quartic coupling, a definitive flagship process has yet to be identified. However, triple Higgs production is considered a strong candidate for future colliders, such as the FCC or a muon collider.

With higher collision energies anticipated in future experiments, high-energy electroweak VBS processes become particularly compelling. At higher energies, the energy-growing behavior of the VBS signal allows it to stand out more distinctly from the background. Both amplitudes and cross sections exhibit stronger energy growth as the multiplicity of final states increases. For instance, as discussed in Sec.~\ref{sec:PUV}, cross sections in $2 \to 4$ VBS processes are expected to grow significantly with energy when the Higgs self-couplings are modified. Similarly, energy growth in $2 \to 3$ or $2 \to 4$ VBS processes arises when the dominant EFT operator involves derivatives.

In this work, we analyze three sets of processes: $V_L V_L \to h^4$, $V_L V_L \to V_L V_L hh$, and $V_L V_L \to V_L^4$. We compute $E_*$, the energy scale at which perturbative unitarity is violated at the tree level, and the cross sections as functions of the collision energy $E_{\rm cm} = \sqrt{\hat{s}}$, considering modifications of the Higgs cubic and quartic couplings, as well as various operators in HEFT and SMEFT. To simplify high-energy calculations, we use the equivalence theorem, which relates longitudinal vector boson states to NG bosons. Calculations with NG bosons make the energy growth in the amplitude and cross section manifest. For comparison, we analyze the relevant $2 \to 3$ VBS processes, which serve as the minimal probes of the Higgs potential, rather than $2 \to 2$ VBS \cite{Chang:2019vez, Falkowski:2019tft, Mahmud:2024iyn}. It is therefore crucial to investigate whether $2 \to 4$ VBS processes can provide even greater sensitivity to the Higgs potential than $2 \to 3$ VBS processes.

Our results show that in the case of Higgs potential modifications, the $E_*$ values for $2 \to 4$ VBS processes are significantly lower than those for the corresponding $2 \to 3$ VBS processes, typically satisfying $E_{*, 2 \to 3} \sim 10 \cdot  E_{*, 2 \to 4}$, as in Eq.~\eqref{eq:3v4}. This finding, presented in Sec.~\ref{sec:Higgs-potential} and Fig.~\ref{fig:res1}, is further analyzed and explained in Secs.~\ref{sec:PUV} and~\ref{sec:x-sec}. Despite the general suppression of high multiplicity signal yields due to the $(E_{\rm cm} / E_*)^{2(n-2)}$ factor, where $n$ the multiplicity of the final state, the much lower $E_{*, 2 \to 4}$ value allows the $2 \to 4$ cross section to eventually exceed that of $2 \to 3$ at sufficiently high energies. Remarkably, this transition occurs at collision energies around $\sqrt{\hat{s}} = {\cal O}(2$–$5)$\,TeV, which are within the reach of future colliders, as demonstrated in Figs.~\ref{fig:potentialh} and \ref{fig:potentialall} in Sec.\ref{sec:x-sec}.

Additionally, while a detailed background analysis is beyond the scope of this work, it is worth noting that the background for the $2 \to 4$ processes could be smaller than that for the $2 \to 3$ processes at high energies. If this is the case, the significance of the $2 \to 4$ processes could surpass that of the $2 \to 3$ processes even before their cross sections become comparable. This potential reduction in background is another advantage of high multiplicity final states in future colliders.
{Ultimately we need to compare $h^n$ final states and final states with more gauge bosons. As the Higgs boson primarily decays into jets, the $V_LV_Lhh$ and $V_L^4$ final states may be easier to probe at colliders. However, the backgrounds associated with the transverse modes would be sizable and irreducible, so comparing final states with different numbers of Higgs bosons requires a thorough background analysis.  }

Having multiple independent processes provides a broader perspective on the Higgs potential. Relying on a single observable offers limited insight into potential BSM effects, whereas multiple observables enable cross-comparisons that reveal richer information. From this perspective, we propose the $2 \to 4$ processes as attractive observables in the high energy regime. For instance, the VBS process with $V_L V_L hh$ final states is sensitive to $\delta_{3,4}$, while the $h^4$ final state depends on $\delta_5$ in addition to $\delta_{3,4}$. Conversely, the SMEFT operator $(H^\dag H)^3$ predicts a specific relationship among $\delta_{3,...,6}$, leading to a partial cancellation in the matrix element. Consequently, the cross section for the $W_L W_L \to h^4$ process does not grow rapidly enough to exceed any of the $2 \to 3$ cross sections, as shown in Fig.~\ref{fig:potentialsmeft}. A similar cancellation occurs in the $W_L W_L \to W_L W_L hh$ process. In contrast, the $W_L W_L \to W_L^4$ process emerges as the most prominent among the $2 \to 4$ VBS processes, a behavior distinct from the scenario where only $\delta_3 \neq 0$.

To expand on this analysis, we also consider several HEFT operators involving derivatives, motivated by various UV models, as discussed in Sec~\ref{sec:HEFTderivative}. For the same sets of VBS processes, we compute the energy scale of unitarity violation and the cross sections. As shown in Fig.~\ref{fig:res2}, we find that, unlike in EFTs without derivatives, the energy scales of unitarity violation in the $V_L V_L \to V_L V_L h h$ and $V_L V_L \to V_L^4$ processes do not decrease with higher multiplicity; instead, $E_{*, {2 \to 3}} \sim E_{*, {2 \to 4}}$. Consequently, the cross sections for the $V_L V_L \to V_L V_L h h$ and $V_L V_L \to V_L^4$ processes do not exceed those of the $V_L V_L \to V_L V_L h$ processes before the EFT becomes invalid although the $V_L V_L \to hhh,\,h^4$ processes behave differently as illustrated in Fig.~\ref{fig:potentialdw}. This highlights the importance of lower $E_*$ values for cross section growth. If $E_*$ values are similar between the $2 \to 3$ and $2 \to 4$ processes, higher energy growth leads to a suppression in the cross section due to the higher powers of $E/E_*$ when ${E} < E_*$.

As shown in this work, the $2 \to 4$ VBS processes warrant a detailed analysis for high-energy colliders beyond the LHC, particularly when including background studies. However, simulating these processes presents significant challenges because $2 \to 4$ VBS is $2 \to 6$ scattering including the incident fermions that emit the high-energy gauge bosons. This makes Monte Carlo simulations highly inefficient. 
{However, previous studies on the $V_LV_L\to hhh$ process at a muon collider~\cite{Stylianou:2023xit, Chiesa:2020awd} have shown that the expected signal yield is sufficient to potentially provide competitive sensitivity. Although a complete analysis including all backgrounds has not yet been performed, some of the backgrounds are considered in Ref.~\cite{Stylianou:2023xit}.} Our work demonstrates that the $2\to 4$ VBS can yield an even larger signal rate if the collision energy is sufficiently high. To improve the computational efficiency, one practical prescription is to treat the gauge bosons as partons, a method {developed by Refs.~\cite{DAWSON198542, KANE1984367}} that has been implemented for the muon collider {by Ref.}~\cite{Ruiz:2021tdt}. Additionally, tagging the polarization of the gauge bosons could help extract signal events more effectively, as the energy-growing behavior is specific to the longitudinal modes.

Finally, one may wonder whether the $2\to 5$ VBS is more useful than the $2\to 4$ VBS. The $E_*$ value of the $2\to 5$ VBS is lower than that of the $2\to 4$ VBS, with $E_{*,2\to 4} \sim 2 \cdot E_{*, 2\to 5}$, as shown in Eq.~\eqref{eq:4v5}. However, we find that the energy scale at which $W_L W_L \to h^5$ and $h^4$ processes have the same cross section is higher than $E_\sigma$, the energy scale where the cross sections of $W_L W_L \to h^4$ and $h^3$ are the same. This indicates that higher collision energies are required for the $2\to 5$ VBS to become most significant. Furthermore, even at the parton level, the $2\to 5$ scattering is more computationally complex.

\section{Acknowledgements}
We thank Maria Mazza for the useful discussion. This work was supported by, in part, the US Department of Energy grant DE-SC0010102.
KT was also supported by JSPS KAKENHI 21H01086.

\appendix
 

\section{Tabulated Values of the $2\to 4$ Processes' Results}\label{app:values24}

In this section, we include the tabulated version of the results in Figs.~(\ref{fig:res1}, \ref{fig:res2}).

\subsection{Potential Modifications}
\begin{table}[htb]
\centering
    \begin{tabular}{||c|c|c|c|c||}
    \hline
    Process &  $ \frac{y^2}{(4 \pi^2)}(H^\dag H)^2 \ln(H^\dag H)$ & $ \kappa_\frac{2}{3}( H^\dag H)^\frac{2}{3}$ & $  \frac{m_h^2}{2v} \delta_3 h^3 + \frac{m_h^2}{8v^2} \delta_4 h^4$ & $\kappa_\frac{1}{2}(H^\dag H)^\frac{1}{2}$\\ 
    \hline 
    $W_L W_L \to h^4$ & 201 & 84.1 & $\frac{1}{\sqrt{|\delta_3-\frac{1}{3} \delta_4|}} \times$20.9 & 50.7  \\
    \hline
    $Z_L Z_L \to h^4$ & 239 & 100& $\frac{1}{\sqrt{|\delta_3-\frac{1}{3} \delta_4|}} \times$24.9 & 60.3  \\
    \hline
    $Z_L Z_L \to Z_L Z_L hh$ & 88.0 & 47.6 & $\frac{1}{\sqrt{|\delta_3-\frac{1}{5} \delta_4|}} \times$14.2 & 31.4 \\
    \hline
    $W_L Z_L \to W_L Z_L hh$ & 108 & 58.3 & $\frac{1}{\sqrt{|\delta_3-\frac{1}{5} \delta_4|}} \times$17.4 & 38.5 \\
    \hline
     $Z_L Z_L \to W_L W_L hh$ & 128 & 69.4 & $\frac{1}{\sqrt{|\delta_3-\frac{1}{5} \delta_4|}} \times$20.7 & 45.7 \\
    \hline
     $W_L W_L \to Z_L Z_L hh$ & 128 & 69.4 & $\frac{1}{\sqrt{|\delta_3-\frac{1}{5} \delta_4|}} \times$20.7 & 45.7  \\
    \hline  
    $W_L W_L \to W_L W_L hh$ & 76.2 & 41.2& $\frac{1}{\sqrt{|\delta_3-\frac{1}{5} \delta_4|}} \times$12.3  & 27.2 \\
    \hline
    \end{tabular}
    \caption{$E_*$ values for the Higgs potential modifications that have been added to the SM with final states involving the Higgs boson, given in TeV. Here, $y=1$, $\kappa_{\frac{2}{3}}\simeq (80$ GeV$)^\frac{8}{3}$, and $\kappa_{\frac{1}{2}}\simeq(80$ GeV$)^3$. A quintic coupling will add a $ \frac{10 v^2}{3 m_h^2} \delta_5$ term inside the absolute value of the  two $h^4$ final state results in column 3. }
    \label{table:res1}
\end{table}

\begin{table}[htb]
\centering
    \begin{tabular}{||c|c|c|c|c||}
    \hline
    Process &  $ \frac{y^2}{(4 \pi^2)}(H^\dag H)^2 \ln(H^\dag H)$ & $ \kappa_\frac{2}{3}( H^\dag H)^\frac{2}{3}$ & $  \frac{m_h^2}{2v} \delta_3 h^3 + \frac{m_h^2}{8v^2} \delta_4 h^4$ & $\kappa_\frac{1}{2}(H^\dag H)^\frac{1}{2}$\\ 
    \hline 
    $W_L W_L \to W_L^4$ & 52.3 & 54.2 & $\frac{1}{\sqrt{|\delta_3|}} \times$18.9 & 37.4  \\
    \hline
    $Z_L Z_L \to Z_L^4$ & 61.6 & 63.8& $\frac{1}{\sqrt{|\delta_3|}} \times$22.2 & 44.0  \\
    \hline
    $Z_L Z_L \to Z_L Z_L W_L W_L$ & 74.0 & 76.7 & $\frac{1}{\sqrt{|\delta_3|}} \times$26.7 & 52.5 \\
    \hline
    $W_L Z_L \to W_L Z_L Z_L Z_L$ & 81.9 & 84.9& $\frac{1}{\sqrt{|\delta_3|}} \times$29.6 & 58.5 \\
    \hline
    $W_L Z_L \to W_L Z_L W_L W_L$ & 76.2& 79.0 & $\frac{1}{\sqrt{|\delta_3|}} \times$27.6 & 54.5 \\
    \hline
     $W_L W_L \to Z_L Z_L W_L W_L$ & 76.2 & 79.0 & $\frac{1}{\sqrt{|\delta_3|}} \times$27.6 & 54.5  \\
    \hline 
     $Z_L Z_L \to W_L^4$ & 108 & 112 & $\frac{1}{\sqrt{|\delta_3|}} \times$39.0 & 77.0\\
    \hline 
    $W_L W_L \to Z_L^4$ & 116 & 120& $\frac{1}{\sqrt{|\delta_3|}} \times$41.8 & 82.8\\
    \hline
    \end{tabular}
    \caption{$E_*$ values for the Higgs potential modifications that have been added to the SM Lagrangian in the vector boson-only final states, given in TeV. Here, $y=1$, $\kappa_{\frac{2}{3}}\simeq (80$ GeV$)^\frac{8}{3}$, and $\kappa_{\frac{1}{2}}\simeq(80$ GeV$)^3$. }
    \label{table:res2}
\end{table}
\newpage
\subsection{Derivative Modifications}

\begin{table}[htb]
\centering
    \begin{tabular}{||c|c|c|c|c||}
    \hline
    Process  & Fermion $(y=1.4) $ & Fermion $(y=2.6)$ & Scalar singlet & Scalar doublet \\ 
    \hline 
    $W_L W_L \to h^4$  & 8.53 & 5.42 & \textit{96.9} & 4.23 \\
    \hline
    $Z_L Z_L \to h^4$  &  9.30 & 5.91 & \textit{115} & 4.61  \\
    \hline
    $Z_L Z_L \to Z_L Z_L hh$  & 18.9 & 11.7 & \textit{41.9} & 10.2 \\
    \hline
    $W_L Z_L \to W_L Z_L hh$ & 5.23 & 4.27 & 8.36 & 4.83 \\
    \hline
     $W_L W_L \to Z_L Z_L hh$ & 4.82 & 3.93 & 7.71  & 4.47  \\
    \hline  
    $W_L W_L \to W_L W_L hh$ & 5.31& 4.34 & 8.43   & 4.86 \\
    \hline
    \end{tabular}
    \caption{$E_*$ values for the HEFTs that have been added to the SM in final states involving the Higgs boson, given in TeV. The values in italics indicate a contribution from $\cal M$$\propto E_{\rm cm}^0 $ from the scalar singlet bare mass vanishing. For the scalar singlet, $\lambda=8.3$, for the scalar doublet, $M= 45$ GeV, where $M = \kappa^2 /\lambda_\Sigma$.}
    \label{table:res1d}
\end{table}

\begin{table}[htb]
\centering
    \begin{tabular}{||c|c|c|c|c||}
    \hline
     Process  & Fermion $(y=1.4) $ & Fermion $(y=2.6)$ & Scalar singlet & Scalar doublet\\ 
    \hline 
    $W_L W_L \to W_L^4$ & 6.91 & 5.08 & 11.5 & 8.16 \\
    \hline
    $Z_L Z_L \to Z_L Z_L W_L W_L$ & 8.04 & 5.91 & 13.3 & 9.49 \\
    \hline
    $W_L Z_L \to W_L Z_L Z_L Z_L$& 7.86 & 5.78  & 13.0 & 9.28 \\
    \hline
    $W_L Z_L \to W_L Z_L W_L W_L$ & 8.07 & 5.94 & 13.4 & 9.52 \\
    \hline
     $W_L W_L \to Z_L Z_L W_L W_L$ & 8.76 & 6.45 & 14.5 &  10.3 \\
    \hline 
     $Z_L Z_L \to W_L^4$ & 6.85 & 5.04 & 11.4  & 8.09\\
    \hline 
    $W_L W_L \to Z_L^4$ & 6.60 & 4.86 & 11.0  & 7.79 \\
    \hline
    \end{tabular}
    \caption{$E_*$ values for the HEFTs that have been added to the SM in vector boson-only final states, given in TeV. For the scalar singlet, $\lambda=8.3$, for the scalar doublet, $M= 45$ GeV, where $M = \kappa^2 /\lambda_\Sigma$.  As a consequence of momentum conservation, the $Z_L Z_L \to Z_L^4$ process has a matrix element with no energy dependence, as in the $2\to3$ case. }
    \label{table:res2d}
\end{table}
\newpage
\section{Tabulated Values of the Lower Multiplicity Processes' Results}\label{app:values23}

In this section, we include the tabulated version of the results in Ref.~\cite{Mahmud:2024iyn}.

\subsection{Potential Modifications}
\begin{table}[htb]
\centering
    \begin{tabular}{||c|c|c|c|c||}
    \hline
    Process &  $ \frac{y^2}{(4 \pi^2)}(H^\dag H)^2 \ln(H^\dag H)$ & $ \kappa_\frac{2}{3}( H^\dag H)^\frac{2}{3}$ & $  \frac{m_h^2}{2v} \delta_3 h^3 + \frac{m_h^2}{8v^2} \delta_4 h^4$ & $\kappa_\frac{1}{2}(H^\dag H)^\frac{1}{2}$\\ 
    \hline 
    $W_L W_L \to hh h$ & 2.7 & 1.7 & $\frac{1}{|\delta_3- \frac{1}{3} \delta_4|} \times$0.12 & 0.68  \\
    \hline
    $Z_L Z_L \to Z_L Z_L h$ & 0.72 & 0.77 & $\frac{1}{|\delta_3|} \times$0.094 & 0.37 \\
    \hline
    $W_L Z_L \to W_L Z_L h$ & 1.1 & 1.2 & $\frac{1}{|\delta_3|} \times$0.14 & 0.55 \\
    \hline
     $Z_L Z_L \to W_L W_L h$ & 1.6 & 1.6 & $\frac{1}{|\delta_3|} \times$0.20 & 0.79 \\
    \hline
     $W_L W_L \to Z_L Z_L h$ & 1.6 & 1.6 & $\frac{1}{|\delta_3|} \times$0.20 & 0.79  \\
    \hline  
    $W_L W_L \to W_L W_L h$ & 0.55 & 0.58& $\frac{1}{|\delta_3|} \times$0.071  & 0.28 \\
    \hline
    \end{tabular}
    \caption{$E_*$ values for the Higgs potential modifications that have been added to the SM in the 3-body final states, given in $10^3$~TeV.  Only the $hhh$ final state has a $\delta_4$ dependence.}
    \label{table:res1old}
\end{table}

\subsection{Derivative Modifications}

\begin{table}[htb]
\centering
    \begin{tabular}{||c|c|c|c|c||}
    \hline
    Process &  Fermion $(y=1.4)$ & Fermion $(y=2.6)$ & Scalar singlet & Scalar doublet \\ 
    \hline 
    $W_L W_L \to hh $ & 25 & 6.9 & - & 9.7  \\
    \hline
     $W_L W_L \to W_L W_L$ & 6.8 & 3.7 & 18 & 9.0 \\
    \hline
     $W_L W_L \to Z_L Z_L$ & 5.7 & 3.1 & 15 & 7.6 \\
    \hline
     $Z_L Z_L \to W_L W_L$ & 5.7 & 3.1 & 15 & 7.6 \\
    \hline  
    $W_L Z_L \to W_L Z_L$& 6.8 & 3.7 & 19  & 9.0 \\
    \hline
    \end{tabular}
    \caption{$E_*$ values for the HEFTs that have been added to the SM given in TeV for the $2\to 2$ processes. For the scalar singlet, $\lambda=8.3$, for the scalar doublet, $M= 45$ GeV, where $M^2= \frac{\kappa^2}{\lambda_\Sigma}$. The $hh$ final state has no unitarity violation scale from the derivative terms when the singlet scalar bare mass is set to zero.}
    \label{table:res22}
\end{table}

\begin{table}[htb]
\centering
    \begin{tabular}{||c|c|c|c|c||}
    \hline
    Process &  Fermion $(y=1.4)$ & Fermion $(y=2.6)$ & Scalar singlet & Scalar doublet \\ 
    \hline 
    $W_L W_L \to hhh $ & 14 & 7.2 & \textit{630} & 6.1  \\
    \hline
     $W_L W_L \to W_L W_Lh$ & 5.2 & 3.8 & 9.9 & 5.3 \\
    \hline
     $W_L W_L \to Z_L Z_Lh$ & 4.7 & 3.4 & 8.9 & 4.8 \\
    \hline
     $Z_L Z_L \to W_L W_Lh$ & 4.6 & 3.4 & 8.9 & 4.7 \\
    \hline  
    $W_L Z_L \to W_L Z_Lh$& 5.2 & 3.8 & 10  & 5.3 \\
    \hline
    \end{tabular}
    \caption{$E_*$ values for the chosen HEFTs that have been added to the SM given in TeV for the $2\to 3$ processes. The values in italics indicate a contribution from $\cal M$$\propto E^0 $ from the scalar singlet bare mass vanishing. For the scalar singlet, $\lambda=8.3$, for the scalar doublet, $M= 45$ GeV, where $M^2= \kappa^2 /\lambda_\Sigma$.}
    \label{table:res23}
\end{table}

\bibliographystyle{JHEP} 
\bibliography{main} 

\end{document}